 \documentclass[preprint]{aastex}

 \def\baselinestretch{1.15}
\newcommand{\myemail}{qryuan@email.njnu.edu.cn}
\newcommand{\kms}{km~s$^{-1}$}

\shorttitle{ BATC and SDSS Photometries of Abell~2255 }

\shortauthors{Yuan et al.}

\begin{document}

\title{MULTICOLOR PHOTOMETRY OF THE GALAXIES IN ABELL~2255 BY THE BATC AND
 SDSS SURVEYS}


\author{\large Qirong Yuan\altaffilmark{1,2},
 Xu Zhou\altaffilmark{2}, and
 Zhaoji Jiang\altaffilmark{2}
% ,\\ Jiansheng Chen\altaffilmark{2},
% Jun Ma\altaffilmark{2}, and Jin Zhu\altaffilmark{2}
}

\altaffiltext{1}{Department of Physics, Nanjing Normal University,
                 NingHai Road 122, Nanjing 210097, China; \myemail}

\altaffiltext{2}{National Astronomical Observatories, Chinese Academy of
Sciences, Beijing 100012, China}
% zhouxu,chen@vega.bac.pku.edu.cn

\begin{abstract}

We present our optical multicolor photometry for the nearby cluster of
galaxies Abell~2255 with 13 intermediate filters in the
Beijing-Arizona-Taiwan-Connecticut (BATC) system which cover an optical
wavelength range from 3000 to 10000 \AA. The spectral energy distributions
(SEDs) in the optical band for more than 7000 sources are achieved down to $V
\sim 20$  in a field of 58' $\times$ 58' centered on this rich cluster.
Abell~2255 has been recently observed by the Sloan Digital Sky Survey (SDSS)
down to $V \sim 17.5$ spectroscopically and $r' \sim 22.0$ photometrically. A
method of combining the SDSS photometric data in five broad bands and the BATC
SEDs is then explored. A sample of 254 galaxies with known redshifts in the
region of Abell~2255 is constructed for testing the reliability of the method
of combining SEDs. Our application of the technique of photometric redshift on
this sample shows that the combined SEDs with higher resolution could lead to
a more accurate estimate of photometric redshift. Based on 214
spectroscopically confirmed member galaxies, spatial and dynamical properties
of this cluster are studied. Bimodality and large dispersion in the velocity
distribution indicate that Abell~2255 is an unrelaxed system. A tight
color-magnitude correlation for 188 known early-type cluster galaxies is
found. After an exclusion of 254 extragalactic sources with known redshifts,
the combined SEDs for 2522 galaxies allow a further membership selection by
the photometric redshift technique and color-magnitude correlation. As a
result, 313 galaxies with their photometric redshifts between 0.068 and 0.090
are selected as new cluster members. The combined SEDs and estimated redshifts
for previously known and newly-selected member galaxies are cataloged. On the
basis of the enlarged sample of member galaxies, the spatial distribution,
localized velocity structure, color-magnitude relation, and luminosity
function of cluster galaxies are discussed. The reverse peculiar velocities
are found for two satellite subclusters located on opposite sides of the
central concentration, which supports an ongoing merger in Abell~2255.

\end{abstract}


%% and \citet commands to identify citations. The citations are

\clearpage

\section{INTRODUCTION}

As the largest gravitationally bound systems in the Universe, galaxy clusters
have been attracting more and more observational efforts in the last decade.
According to the hierarchical models for large-scale structure formation in
the picture of cold dark matter, galaxy clusters are built up by the process
of hierarchical clustering in which larger clusters are formed by the
coalescence of smaller clusters. Strong gravitational encounters between
galaxy clusters are essential in forming the structures we observe today. This
scenario is supported by the recent X-ray observations of some nearby rich
clusters which give convincing evidence for a ongoing merger event (Colless \&
Dunn 1996). The rich cluster of galaxies, Abell~2255, is one of the
well-studied examples.

The nearby galaxy cluster Abell~2255 has recently been observed in radio,
optical, and X-ray bands, and a complicated picture of its dynamics, structure
of the member galaxies, and intergalactic medium is unveiled. From its radio
maps at 20 cm, a diffuse halo source appears at the cluster center, and 6
tailed radio sources are found to be associated with cluster galaxies (Feretti
et al 1997). The electrons can be accelerated throughout the cluster to power
a radio halo in the outer region if this cluster is undergoing a merger
(Tribble 1993). The asymmetric structure in Abell~2255 detected by ROSAT
imaging observations and the nonisothermal temperature map of the intracluster
gas in this cluster support the discovery of a currently undergoing merger
(Davis \& White 1998). From the optical point of view, there are two
comparably bright galaxies near the cluster center, which is thought to be the
evidence of recent merger events with two less massive systems of galaxies
(Bird 1994). This scenario is also supported by the unusually large radial
velocity dispersion of $\sim$ 1221 \kms, which may be indicative of the
dynamically unrelaxed phase (Zabludoff et al. 1990). Furthermore, the optical
images show that the X-ray emission centroid does not coincide with any bright
galaxy or galaxy subgroup (Burns et at. 1995). These intriguing observational
features may tell us that this cluster is not a simple relaxed structure, but
is still forming at the present epoch.

However, no obvious evidences of spatial or kinematical substructure in
Abell~2255 was found in optical band (Stauffer et al. 1979; Zabludoff et al.
1990). It is mainly because only the limited number of cluster galaxies
brighter than V $\sim$ 16.0 have been spectroscopically observed a decade ago.
Recently, accurate photometric data for the objects in this cluster were
obtained and distributed by the Sloan Digital Sky Survey (SDSS). However, the
SDSS spectroscopic survey covers merely the galaxies brighter than $V \sim
18.0$. Based on the luminosity function of cluster galaxies (Lobo, et al.
1997), a typical galaxy with $M_V \sim -18.0$ has an apparent magnitude of $V
\sim 20.5$, fainter than the limit of the SDSS spectroscopic survey. For a
better understanding of structure and dynamics of this cluster, the faint
member galaxies ($18.0 < m_v < 21.5$) should be taken into account. This paper
will present our optical photometry of the galaxies in Abell~2255 with the
Beijing-Arizona-Taiwan-Connecticut (hereafter BATC) multicolor system. The
BATC multicolor photometric survey aims mainly at obtaining the spectral
energy distribution (SED) information of galaxies with redshift less than 0.5
(Xia et al. 2002), by using the 60/90 cm Schmidt Telescope of Beijing
Astronomical Observatory (BAO) with 15 intermediate-band filters. Our previous
work shows that the multicolor photometry with 1-meter class Schmidt Telescope
can provide the SED information for all the objects within a large field of
view ($\sim$ 1 deg$^2$) centered on a nearby galaxy cluster, which will
facilitate the follow-up investigations on membership selection and the
color-magnitude relation (Yuan et al. 2001). Since the accuracy of photometric
redshifts is heavily dependent upon the number of filters (i.e., spectral
resolution) and the photometric accuracy in each band (Xia et al. 2002), the
SDSS photometric data in five photometric bands (Fukugita et al. 1996) will be
combined with the BATC SEDs for all the objects detected within our field.
Current study will result in an SED catalog of not only previously known
(bright) member galaxies but the newly-selected faint members. Based on the
enlarged sample of cluster galaxies, the spatial distribution, dynamics, color
properties and luminosity function of cluster galaxies in Abell~2255 can be
then addressed.

The structure of this paper is as follows. In \S 2, we present the BATC
multicolor photometric observations and data reduction, as well as the
cross-identification between the BATC and SDSS photometric catalogs. The
methodology of combining the SDSS SEDs with the BATC SEDs are detailed in \S
3.
The analysis of SED features for the previously known member galaxies is
shown
in \S 4, and the combined SEDs of known galaxies in Abell~2255 are also
cataloged. The further SED selection of faint members is given in \S 5. The
dynamical and color properties of the enlarged sample of cluster
galaxies is presented in \S 6, as well as the luminosity function of cluster
galaxies. Finally, we give a summary in \S 7. The cosmological parameters
$H_0$=50 km s$^{-1}$ Mpc$^{-1}$ and $q_0$=0.5 are assumed throughout this paper.

\section{OBSERVATIONS AND DATA REDUCTION}

Abell~2255 is a nearby ($z \sim$ 0.0806, giving a distance modulus of 38.3;
Struble \& Rood 1999) cluster of galaxies with richness class 2 in Abell
(1958), classified as a type II-III cluster by Bautz \& Morgan (1970). The
BATC photometric observations of Abell~2255 were taken with the 60/90 cm f/3
Schmidt Telescope of BAO, located at the Xinglong site with an altitude of
900m. A Ford 2048 $\times$ 2048 CCD camera was equipped at the prime focus of
the telescope. The field of view was $\sim$ 1.0 square degree, and the spatial
scale was $1.7''$ per pixel. The combined image in each filter covers a sky
region of 59 $\times$ 59 arcmin$^2$, defined by a right ascension range from
$17^h 08^m 03^s$.7 to $17^h 16^m 58^s$.6, and a declination range from
$63^{\circ} 36' 31''$.5 to $64^{\circ} 35' 28''$.8 (for equinox of 2000.0).
The BATC filter system includes 15 intermediate-band filters, namely, $a-k$,
$m-p$, covering the whole optical wavelength range from $\sim$ 3000 to 10000
\AA. The details of the BAO Schmidt Telescope, CCD camera and data-acquisition
system can be found elsewhere (Fan et al. 1996; Yan et al. 2000). For
distinguishing explicitly between the SDSS and BATC filter names, in this
paper we refer the SDSS filters and magnitudes to as $u'$, $g'$, $r'$, $i'$,
and $z'$, which corresponds to central wavelengths of 3560, 4680, 6180, 7500
and 8870 \AA. The transmissions of the BATC and SDSS filters can been seen in
Fig. 1.

\smallskip
\centerline{\framebox[12cm][c]{Fig. 1: The transmissions of the BATC and SDSS
filters}}
\smallskip

Only 13 BATC filters were used for the multicolor imaging observations of
Abell~2255. In total, we have made more than 40 $hr$ of exposures. After a
check of the image quality, 113 images with more than 36 $hr$ of exposure were
selected to be combined. The parameters of the BATC filters we used and the
observational details are given in Table 1. The bias subtractions and dome
flat-field corrections were done on the CCD images, using {\em Pipeline 1},
the automatic data reduction code for the BATC survey. The cosmic ray and bad
pixel effect were corrected by comparing the images. Before combination, the
images were re-centered and position calibration was performed by using the
Guide Star Catalog (GSC). The processes of the point spread function (PSF)
fitting and the aperture photometries with different apertures were performed
for the objects detected in the images of at least 3 filter bands, using the
code named as {\em Pipeline 2} (Zhou, et al. 2003).  As a result, the list of
PSF and aperture magnitudes in various filter bands was obtained for all the
objects detected. For the bright galaxies in the region of Abell~2255, we
adopted the aperture magnitudes defined by a fixed aperture with a radius of 4
pixels (i.e., 6.8 arcsec) which is large enough to make different seeing
effects negligible. Although the magnitudes measured with a fixed aperture are
not the same as the total magnitudes of galaxies in the literature, this is a
proper way to obtain the reliable color indices (i.e., relative SEDs) of all
the objects. Since the adopted aperture size is rather large, contamination of
the light by nearby objects may sometimes not be negligible. The number of
sources detected in the combined images are shown in Table 1, along with the
details of the observations.

%table 1
\begin{table}[ht]
\caption[]{The details of the BATC filters and our observations}
\vspace {0.5cm}
\begin{tabular}{ccccccccc}   \hline   \hline
\noalign{\smallskip}  
 No. & Filter & $\lambda_{eff}$ & FWHM &
Exposure &  Number of & Seeing$^a$  & Objects & Limiting \\
   &  name & (\AA) & (\AA) & 
(second) & Images     & (arcsec)  & Detected & mag \\
\noalign{\smallskip}   \hline \noalign{\smallskip} 
  1  & b  & 3907 & 291 & 10800  & 9&  4.74   & 5678 & 20.5 \\
  2  & d  & 4540 & 332 & 13200  &11&  5.45   & 6816 & 20.5 \\
  3  & e  & 4925 & 374 & 12000  &10&  4.32   & 7227 & 20.0 \\
  4  & f  & 5267 & 344 & 11400  &10&  4.98   & 7253 & 20.0 \\
  5  & g  & 5790 & 289 & 7200   & 6&  3.96   & 7337 & 20.0 \\
  6  & h  & 6074 & 308 & 6950   & 7&  4.18   & 7319 & 20.0 \\
  7  & i  & 6656 & 491 & 12900  &12&  4.61   & 6884 & 19.5 \\
  8  & j  & 7057 & 238 & 4800   & 4&  4.36   & 7437 & 20.0 \\
  9  & k  & 7546 & 192 & 4800   & 4&  4.62   & 7386 & 19.0 \\
  10 & m  & 8023 & 255 &14400   &12&  4.11   & 7368 & 19.0 \\
  11 & n  & 8484 & 167 & 9600   & 8&  5.13   & 7047 & 19.0 \\
  12 & o  & 9182 & 247 &12000   &10&  3.81   & 7355 & 18.5 \\
  13 & p  & 9739 & 275 & 12000  &10&  4.21   & 6796 & 18.5 \\
\noalign{\smallskip}   \hline  
\end{tabular}
\end{table}
\vskip -3mm   
\noindent $^a$  This column lists the seeings of
the combined images.

%\smallskip
%\centerline{\framebox[8cm][c]{Table 1: BATC observation log}}
%\smallskip

To obtain the {\em relative} SEDs of objects in this field, the model
calibration was made for the combined images. This method was developed to
calibrate the SEDs of objects in a large field of view on the basis of the SED
library, especially for the BATC multicolor photometric system (Zhou et al.
1999). No calibration images of the standard stars are needed to derive the
relative SEDs.

It can be expected that the SDSS photometric data can be used to do flux
calibration of the BATC SEDs, which will be described in next section.
Therefore we cross-identified all the sources detected by both photometric
observations. All the sources in the BATC catalog within the searching circle
(defined by a radius of 2.0 arcsec) centered on the SDSS sources were
extracted. By comparing the SED features, the identification was rather
unambiguous. According to the morphology classification given in the SDSS
photometric catalog, 3651 point sources (stars) and 2790 extended sources
(galaxies) were found in both surveys. Furthermore, we extracted the BATC and
SDSS photometric information for all known extragalactic objects listed in the
NASA/IPAC Extragalactic Database (NED). There are 340 known sources in our
observation field, among which 274 (81\%) are galaxies and 14 (4\%) are
quasars. The remainder are sources that appear only in one of surveys from
radio, infrared, X-ray, and $\gamma$-ray wavebands. Some 254 galaxies have
been found to have NED-given spectroscopic redshifts.  The counterparts for
those 254 galaxies with known spectroscopic redshifts were found in both of
the BATC and SDSS imaging observations, which offers a good sample for our
further analyses.

\section{METHOD OF COMBINING THE SDSS AND BATC SEDS}

It is clear that combination of the BATC SEDs with the SDSS photometric data
will lead to a more accurate estimate of photometric redshift. Therefore we
tried to derive the SDSS colors of galaxies through the same aperture as
defined by the BATC photometry (i.e., aperture radius $r_{ap} = 4 \times 1.7
=$ 6.8 arcsec), using the photometric parameters provided by the SDSS Early
Data Release (EDR; Stoughton et al. 2002). The SDSS imaging observations of
Abell~2255 were carried out by drift scanning with a spatial scale of 0.4
arcsec pix$^{-1}$ in five broad bandpasses. The effective integration time was
54 seconds, and all images were reduced with software specially designed for
the SDSS data (Lupton et al. 2001).

\subsection{\sl Aperture Correction of the SDSS Photometries for Galaxies}

For a given bright galaxy detected by the SDSS imaging observation, the
observed profile of surface brightness is quantified by some global
photometric parameters, such as the likelihood parameters for various models,
the effective radius ($r_e$) along the major axis, and the {\em model}
magnitude ($m_{model}$) which represents the total magnitude estimated by the
optimum model in the $r'$ band. The three following models were used to fit
the observed surface-brightness profile: (i) the point spread function (PSF)
model, (ii) de Vaucouleurs (1948) $r^{1/4}$ model, and (iii) exponential
model. The model magnitudes in other four bands were calculated using the
effective radius defined by the preferential modelling in the $r'$ band. Some
studies show that the de Vaucouleurs model appears to be a very good fit to
the elliptical galaxies' surface-brightness (de Vaucouleurs \& Capaccioli
1979; Capaccioli et al. 1990). The difference between the {\em model}
magnitude ($m_{model}$) and the magnitude within a given aperture ($m_{ap}$)
can be
derived by %
\begin{eqnarray} \Delta m = m_{ap} - m_{model} = -2.5 \, log
\frac{\int_0^{r_{ap}} 2 \pi r\, I(r) dr} {\int_0^{\infty}2\pi r\, I(r) dr},
\end{eqnarray}
where $I(r)$ is the profile function of surface intensity defined as
\begin{eqnarray}
\begin{array}{ll}
I(r)= & \left\{\begin{array}{ll}
 I_0\, exp\{-7.67[(r/r_e)^{1/4}]\}, &  {\rm as }\,\, 0 <r \le 7r_e, \\
  a_0\, I_0 (8-r/r_e), & {\rm as }\,\, 7r_e <r <8r_e.
\end{array} \right.
\end{array}
\end{eqnarray}
for the de Vaucouleurs $r^{1/4}$ model, or
\begin{eqnarray}
\begin{array}{ll}
I(r)= & \left\{\begin{array}{ll}
      I_0\, exp(-1.68\,r/r_e), & {\rm as } \,\,0 <r \le 3r_e, \\
      b_0\, I_0 (4-r/r_e), & {\rm as } \,\,3r_e <r <4r_e.
\end{array} \right.
\end{array}
\end{eqnarray}
for the exponential model, where the dimensionless quantities $a_0$ and
$b_0$ are intensities relative to the amplitude $I_0$ at $7r_e$ and $3r_e$,
respectively: $a_0=3.8178 \times 10^{-6}$ and $b_0=6.4737 \times 10^{-3}$.
The aperture correction $\Delta m$ is the same for the {\em model}
magnitudes in five SDSS filters, since the model with higher likelihood in
the $r'$ filter is selected to be used in the other bands with the same
effective radius $r_e$ for a specific object, allowing only the amplitude
$I_0$ to vary.

It is well known that the vast majority of stars detected in the BATC
observations have the brightness profiles which can be well modelled by the
point spread function (PSF). Therefore we adopted the PSF magnitudes in the
BATC and SDSS observations as optimal measures of the fluxes of stars.

\subsection{\sl Flux Calibration of the BATC SEDs}

The SDSS model magnitudes are flux-calibrated by comparison with a set of
overlapping standard-star fields calibrated with a 0.5-m ``photometric
Telescope''. The uncertainties of the model magnitudes are also listed in
the SDSS EDR photometric catalog. In general, the error for $u'$ model
magnitude is larger than those in other bands, and brighter sources have
less fitting uncertainties. For a known galaxy having $u' < 19.5 $, the
typical $u'$ error is less than 0.1.

As we mentioned above, the BATC SEDs were calibrated by comparing with the SED
library of bright stars in the BATC field. After the model calibrations, the
relative SEDs for all the objects in the BATC photometric field were
obtained. The zero-point of the BATC SEDs of galaxies can be determined by the
aperture-corrected SDSS SEDs.  Fig. 2 shows the relative SEDs in the BATC and
SDSS multicolor systems for SDSS J171236.07+640508.0, the central early-type
galaxy in Abell~2255 cluster. The zero-point of the BATC SEDs can be derived
simply by averaging the magnitude differences at 6166 and 7480 \AA, the
effective wavelengths of $r'$ and $i'$ filters. The interpolation can be used
to derive the BATC magnitudes at 6166 and 7480 \AA, based on the magnitudes in
the neighboring BATC bands. The flux zero-points for different galaxies are
slightly different, since the deviation of the surface-brightness profile from
the preferential model is different from source to source.

In calculation of flux zero-point for each star, we directly used the PSF
magnitudes measured in BATC and SDSS images. No aperture corrections are
needed for the SDSS magnitudes of stars. The average value of magnitude
differences at 6166 and 7480 \AA\ is also defined as the flux zero-point for
each star. We present the zero-point distribution for all the stars and
galaxies detected by both multicolor surveys in Fig. 3. It can be seen that
the stars have more concentrated distribution of SED zero-points (thick line),
with a peak at 3.24. A wider range of fitting deviations for galaxies leads to
a comparatively scattered zero-point distribution (thin line) with the same
typical zero-point.

\smallskip
\centerline{\framebox[12cm][c]{Fig. 2: BATC and SDSS SEDs for  SDSS
J171236.07+640508.0}} \smallskip

\smallskip
\centerline{\framebox[12cm][c]{Fig. 3: Zero-point distribution}}
\smallskip

As a result, we achieved the combined SEDs for 2790 galaxies and 3651 stars
detected by BATC and SDSS photometry, including the flux-calibrated BATC SEDs
in 13 filter bands and the aperture-corrected SDSS SEDs in 5 filter bands.
This SED catalog can be electronically provided upon request.

To describe the errors of BATC magnitudes, we separated all the objects into
several sub-groups with magnitude interval of 0.5 mag, and the mean
measurement errors at specified magnitudes were derived. We found that the
measurement error for each filter band tends to be larger at fainter depth.
Except for the $n$, $o$ and $p$ magnitudes, the errors are about $0.02$ for
bright stars (say, $m<16.5$), and less than $0.05$ at $m=19.0$. The
measurement errors in $n$, $o$ and $p$ magnitudes are found to be larger
just because of the low sensitivity of our CCD detector in redder filter
bands.

\section{ANALYSES OF 254 GALAXIES WITH KNOWN SPECTROSCOPIC REDSHIFTS}

\subsection{\sl Velocity Distribution and SED Catalog}

Burns et al. (1995) studied the distribution of measured galaxy velocities
for Abell~2255 by using the $ROSTAT$ software.  Two resistant and robust
estimators analogous to the velocity mean and standard deviation, namely the
biweight location ($C_{BI}$) and scale ($S_{BI}$), are defined to
characterize the shape of the velocity distribution (Beers et al. 1990).
Burns et al. (1995) found $C_{BI}=24330^{+203}_{-265}$ \kms and
$S_{BI}=1240^{+203}_{-129}$ \kms with only 39 member galaxies. We have 254
galaxies with known spectroscopic redshifts within our field. The
distribution of spectroscopic redshifts for all these known galaxies is shown
in panel (a) of Fig. 4. There are 214 galaxies occurring in the redshift
range of 21,000 \kms $< cz <$ 29,000 \kms, with a sharp peak at $\sim$ 24,800
\kms. We also calculated the biweight location and scale for these 214
galaxies using the ROSTAT software, and achieved $C_{BI}=24025\pm89$ and
$S_{BI}=1315 \pm 86$ in \kms. Compared with the results in Burns et al.
(1995), the biweight location is smaller and the biweight scale becomes
larger with smaller uncertainties. We fitted the velocity distribution in
Fig. 4b with single Gaussian, but the Kolmogorov-Smirnov test and
Shapiro-Wilk W-test reject a Gaussian distribution at more than 99\%
confidence level.

We tried to detect and quantify the presence of bimodality in velocity
distribution with the KMM algorithm code (Ashman, Bird \& Zepf 1994). Two
equal-variance Gaussian components with velocity mean values of 23942 and
27454 \kms were found to have a good fit with a mixing proportion of
94.2\%:5.8\%. The KMM algorithm also allows partitioning of data into
individual groups. Only 12 (5.8\%) cluster galaxies belong to the Gaussian
component centered at 27454 \kms, and they are scattered in this field. The
confidence level of rejecting the single Gaussian model is nearly 100\%. The
significant deviation from single Gaussian distribution and unusually large
velocity dispersion might reflect some clues of an unrelaxed system.

\smallskip
\centerline{\framebox[12cm][c]{Fig. 4: Radial velocity distribution
for 214 member galaxies}}
\smallskip

Table 2 presents the combined SEDs of 254 galaxies with known spectroscopic
redshifts, which is structured as follows:

\begin{tabular}{ll}
Column 1: & Number of the galaxies (sorted by the distance to cluster
center);\\
Column 2: & R.A. in 2000 epoch, in ``hhmmss.s'' mode, given by NED;\\
Column 3: & Declination in 2000 epoch, in ``ddmmss'' mode, given by NED; \\
Column 4: & Spectroscopic redshift, given by NED;\\
Column 5: & Photometric redshift, estimated by the combined SED;\\
Column 6 - 18: & Photometric magnitudes in 13 BATC filter bands.
The value of 0.0 \\
& means out of field;\\
Column 19 - 23: & Aperture-corrected photometric magnitudes in five SDSS filter
bands.
\end{tabular}

It should be noted that the photometric magnitudes given in Table 2 might be
somewhat different from the magnitudes given in the literature. What we
attempt to obtain is the {\em relative} SEDs within a fixed aperture for the
galaxies in the region of Abell~2255.

\smallskip
\centerline{\framebox[10cm][c]{Table 2: Combined SEDs of 254 known galaxies}}
\smallskip

\subsection{\sl Spatial Distribution and Localized Velocity Structure}

Fig. 5a presents the spatial distribution of 214 known member galaxies within
our field of view, including 168 (88\%) early-type galaxies and 46 (12\%)
late-type galaxies, with respect to the NED-given central position of
Abell~2255, R.A.=17$^h12^m31^s$ and Decl.=64$^{\circ}05'33''$ (for the J2000
equinox). Early- and late-type galaxies are denoted by open circles and
crosses, respectively. It can easily be seen that the early-type galaxies
dominate in the central region. We superpose the contour map of the surface
density which has been smoothed by a Gaussian window with $\sigma = 2$ arcmin.
The spatial distribution of 214 bright member galaxies seems to deviate from
spherically symmetric. The contours of surface density are elongated in an
east-west direction in the central region. The isophotes are offset from the
positions of two cD galaxies (marked as ``$+$''), but seem to have a centroid
located between the cD galaxies and the X-ray emission centroid
(R.A.=17$^h$12$^m$45$^s$ and Decl.=64$^{\circ}$03'54'', mark as ``$\times$'')
derived by the ROSAT imaging observations (Feretti et al. 1997). The contours
show that there might be five satellite clumps (i.e., subclusters) of galaxies
whose surface densities is more than 0.15 galaxies arcmin$^{-2}$ (namely A, B,
C, D, and E) surrounding the main concentration. The contour curves at 0.15
galaxies arcmin$^{-2}$ can be used to define the clump areas.

However, these clumps might be enhancements simply due to projection effects.
The proper way to find out the true substructures is to observe how
significant is the localized variation at clumps' positions in the
line-of-sight velocity distribution. Dressler \& Shectman (1988) designed a
statistical test, so-called $\Delta$-test, for computing the deviation of
local velocity mean and dispersion from those of the overall velocity
distribution. In the spirit of $\Delta$-test, Colless \& Dunn (1996) developed
a new test (so-called $\kappa$-test) for quantifying localized variation in
velocity distribution, by defining a new test statistic ${\kappa}_n$ to
characterize the local deviation on the scale of groups of $n$ nearest
neighbors. The larger ${\kappa}_n$, the greater the possibility that the local
velocity distribution differs from the overall distribution. The probability
that $\kappa_n$ is larger than the observed value $\kappa_n^{obs}$,
$P(\kappa_n>\kappa_n^{obs})$, can be estimated by Monte Carlo simulations by
randomly shuffling velocities. Table 3 gives the results of $\kappa$-test for
214 known member galaxies, and the number of simulations is $10^3$ for all
cases.

%\begin{center}
%table 3
\begin{table}[ht]
%\caption[]
\noindent {Table 3: Result of $\kappa$-Test for 214 Member Galaxies}

\vspace{2mm}
\begin{tabular}{ccccccccc}   \hline   \hline
\noalign{\smallskip}  
 $n$ & 
5      &       6&      7&      8&      9&     12&     15&     20\\
\noalign{\smallskip} \hline \noalign{\smallskip}
$P(\kappa_n>\kappa_n^{obs})$ & 
0.003  &  0.002 &  0.003&  0.006&  0.007&  0.005&  0.005&  0.023\\
\noalign{\smallskip}   \hline  
\end{tabular}
\end{table}

%\end{center}

The presence of substructures in Abell~2255 is strongly supported by the local
velocity variation with a wide range of neighbor sizes. The six nearest
neighbors is the optimum scale on which substructure is most obvious. The
bubble plot in Fig. 5b shows the location of localized variation using
neighbor size $n=6$, and the bubble size for each galaxy is proportional to
$-log[P_{KS}(D>D_{obs})]$. Therefore larger bubbles indicate a greater
difference between local and overall velocity distributions. A close
comparison between two panels in Fig. 5 shows that the bubble clustering
appears at the positions of clumps A, B and C. The clumps D and E seems to be
unreal substructures in the line-of-sight velocity distribution. The dynamics
for substructures A, B and C will be discussed in \S 6.

\smallskip
\centerline{\framebox[12cm][c]{Fig. 5: (a) Spatial distribution for 214
galaxies; (b) Bubble plot}}
\smallskip

\subsection{\sl Color-Magnitude Correlation}

A correlation between color and absolute magnitude for early-type galaxies,
the so-called C-M correlation, has been found for some rich galaxy clusters
(see Bower et al. 1992, and references therein). For the early-type galaxies
in a cluster, the fainter galaxies tend to have colors bluer than the brighter
galaxies do. The measured magnitudes in BATC and SDSS photometries for the
galaxies in Abell~2255 can be directly used to demonstrate the C-M
correlation.

Fig. 6 gives plots of two color indices (C.I.'s), $u'-p$ and $u'-o$, versus
magnitude in the $h$ bandpass for 214 known member galaxies in Abell~2255. The
galaxies with different morphological types are denoted by different symbols.
A very clear C-M correlation can be found for 137 elliptical galaxies, in the
sense that the brighter galaxies are redder. We have following linear fits by
method of least squares:
$$u'-p = -0.136(\pm0.04)h + 5.758(\pm0.669),$$
$$u'-o = -0.134(\pm0.04)h + 5.606(\pm0.638).$$
The linear fits are plotted as solid lines, and the dashed lines in panels (a)
and (b) represent $\sigma_{C.I.} = 0.669 $ and 0.638, respectively. More than
95\% elliptical galaxies are distributed between the dashed lines. The linear
relations are quite similar to those we found for bright galaxies in Abell
2634 (Yuan et al. 2001). It is evident that spiral galaxies do not obey such a
correlation, and lenticular galaxies behave as something between the
ellipticals and the spirals in the C-M relation. Based on the significantly
different behavior for galaxies with different morphologic types, the C-M
diagram can be used to distinguish early- and late-type galaxies (e.g.,
Colless \& Dunn 1996).

\smallskip
\centerline{\framebox[12cm][c]{Fig. 6: Color-magnitude diagram for 214 known
member galaxies}} \smallskip

\subsection{\sl Application of the Photometric Redshift Technique}

The technique of photometric redshift can be used to estimate the redshifts of
galaxies with the combined SED information. For a given object, the
photometric redshift, $z_{phot}$, corresponds to the best fit (in the
$\chi^2$-sense) of its photometric SED by the set of template spectra. Based on
the standard SED-fitting code called HYPERZ (Bolzonella, Miralles, \& Pell\'o
2000), the procedures for estimating the photometric redshifts have been
developed especially for the BATC multicolor photometric system (Xia et al.
2002). This technique is traditionally used to search for galaxies or AGNs with
comparatively high redshifts. Thanks to the large number of filters (13 BATC
bands plus 5 SDSS bands), the accuracy of photometric redshift estimates can be
expected to be largely improved even for nearby galaxies. The sample of 254
galaxies with known spectroscopic redshifts allows an opportunity to check
whether the method of SED combination for the SDSS and BATC surveys can lead to
a more accurate $z_{phot}$ estimate for the nearby galaxies with $z < 0.5$.

Fig. 7 shows the plot of photometric redshift ($z_{phot}$) versus
spectroscopic redshift ($z_{sp}$) for 254 galaxies, including the member
galaxies and background galaxies. In our calculations, only the normal
galaxies were taken into account in the reference templates, and the reddening
law with a $A_V \sim 0.3$ by Calzetti et al. (2000) was adopted. The
photometric redshift for each known galaxy was searched within a redshift
range from 0.0 to 1.0, with a searching step of 0.01. The dotted line in Fig.
7 corresponds to $z_{phot}=z_{sp}$, and the error bar in the $z_{phot}$
determination at 68\% confidence level is also given. It can easily be seen
that our $z_{phot}$ estimate is basically consistent with the spectroscopic
redshift: more than 90\% member galaxies in Abell~2255 are found to have
$z_{phot} \sim 0.075 \pm 0.015$, and the background galaxies with $z_{sp} <
0.5$ also scattered along the critical line $z_{phot}=z_{sp}$. Compared with
our previous $z_{phot}$ estimate for the nearby galaxies in Abell 2634 ($z
\sim 0.03$) on the basis of the BATC SED information only (see Fig. 6 in Yuan
et al. 2001), this run of $z_{phot}$ determination has smaller uncertainties
and deviations in general. Fig. 8 shows an example for the best fit to the
combined SED of the early-type galaxy SDSS J171236.07+640508.0. The combined
SED has a higher spectral resolution, which is crucial for decreasing the
uncertainty of $z_{phot}$ estimate. This result demonstrates not only the
efficiency of our SED-fitting procedures, but also the reliability of our
method of combining the photometric data from the SDSS and BATC photometric
observations.

\vskip 5mm
%-----------------------------------------------------------------
\centerline{\framebox[12cm][c]{Fig. 7: $z_{sp} - z_{phot}$ for 254 known
galaxies}}

\vskip 5mm

%---------------------------------------------------------------------
\centerline{\framebox[12cm][c]{Fig. 8: The best fit of SDSS
J171236.07+640508.0}}

Our method described above opens a straightforward way for obtaining the
optical SEDs from the SDSS and BATC multicolor photometries of the faint
galaxies. Our investigations on spatial distribution and color-magnitude
relation of the galaxies in Abell~2255 can then be extended to an
unprecedented depth. It should be noted that the error in the photometric
redshift is much larger than the cluster velocity dispersion.

\section{SED SELECTION OF FAINT CLUSTER GALAXIES}

In total, 6441 objects within our BATC observation field are found to have the
counterparts in the SDSS imaging observations, of which 2776 objects are
characterized by the SDSS star/galaxy separation pipeline, {\em frames}, as
extended sources (i.e., galaxies). After an exclusion of 254 known galaxies,
there remain 2522 galaxies without any redshift information. Based on their
combined SEDs, we applied the photometric redshift technique to them, and
derived their optimal estimates of $z_{phot}$ values.

Since the error in $z_{phot}$ determination is larger than the
intrinsic dispersion (say, biweight scale $S_{BI}$) in distribution of
spectroscopic redshift, the $z_{phot}$ distribution for faint cluster galaxies
should peak at $\sim$ 0.08 with a larger dispersion. Fig. 9 shows the $z_{phot}$
distribution for 2522 galaxies. The peak at $z_{phot} \sim$
0.08 should be associated with Abell 2255. Based on the fact that more than 98\%
known member galaxies are found to have their redshifts within a range of 20,400
\kms $< cz <$ 27,000 \kms (see Fig. 4), we conservatively adopt this range as
the limits for membership determination. As a result, 328 galaxies having their
photometric redshifts between 0.068 and 0.090 are selected to be member
candidates.

\vskip 5mm
\centerline{\framebox[12cm][c]{Fig. 9: The $z_{phot}$ distribution for
2522 galaxies}}

Among 328 member candidates, 266 (81\%) galaxies are regarded as early-type
galaxies by our SED-fitting procedures. A further selection by C-M correlation
for these early-type galaxies can be performed. We find that the majority of
early-type candidates agree with the C-M relation derived by 137 bright
early-type galaxies in \S4.3. However, there are 48 early-type candidates with
colors beyond the 2$\sigma$ deviation of intercept, and they might have been
misclassified. Taking only the templates of late-type galaxies, we performed
the photometric redshift estimate again for these 48 candidates, and 15
candidates are found to have photometric redshifts beyond the range of 0.068
$< z_{phot} <$ 0.090.  They are therefore regarded as non-member galaxies, and
the remaining 33 galaxies are regarded as late-type member galaxies.

Finally, 313 faint galaxies (including 218 early-types and 95 late-types)
are selected as faint member galaxies in Abell~2255.  Table 4 presents the
catalog of SED information for these new members, as well as SDSS-measured
position, $z_{phot}$ value, and morphological class $T$ (E, S0, Sa, Sb, Sc,
Sd, and Im galaxies are represented by 1 to 7, respectively).

\smallskip
\centerline{\framebox[10cm][c]{Table 4: SED Catalog of 313 galaxies}}
\smallskip

\section{ANALYSES OF THE ENLARGED SAMPLE OF CLUSTER GALAXIES}

\subsection{\sl Velocity Distributions and Substructures}

Some interesting and unusual characteristics in radio, and X-ray bands support
a picture that Abell~2255 may be in the process of merging with a galaxy
cluster/group (Feretti et al. 1997). Fig. 10 shows the redshift distributions
for (i) 203 galaxies in a central region with a radius of 10 arcmin, (ii) 406
early-type galaxies, (iii) 121 late-type galaxies, and (iv) 527 cluster
galaxies in the enlarged sample. Since two peaks, respectively, at 24900 and
23700 \kms can be found in subset (i),(ii) and the total sample, we assumed
two homoscedastic Gaussian components in the KMM algorithm code to search for
bimodality in above velocity distributions. However, none of these tests
indicate that a bimodal distribution is a statistically significant
improvement over the single Gaussian. Compared with the velocity distribution
of 214 known galaxies in Fig.4b, the enlarged sample has a larger biweight
scale (1848 \kms). This should be due to the large error in photometric
redshift estimate for 313 newly-selected member galaxies, and it will probably
smooth the bimodality detected in Fig. 4b.

\centerline{\framebox[12cm][c]{Fig. 10: Velocity distributions for four
samples}}

The vast majority of 313 newly-selected member galaxies are fainter than $h
\sim $17.6, and the early-type faint galaxies dominate in the central
10-arcmin region. Compared with the sample of 214 known cluster galaxies,
these faint galaxies have a higher fraction of late-type galaxies (30\%), and
most of late-type galaxies are scattered around the main concentration. The
spatial distribution of the enlarged sample of 527 cluster galaxies is shown
Fig. 11a. In general the contour map of this rich cluster appears to be
elongated along southwest and northwest directions. The cluster centroid seems
to agree well with that shown in Fig.5a.

\vskip 5mm
\centerline{\framebox[12cm][c]{Fig. 11: (a) Spatial distribution
of 517 galaxies; (b) Bubble plot}}

To observe the substructures in the velocity distribution, we applied the
$\kappa$-test on the enlarged sample of cluster galaxies. The $\kappa$-test
results for $10^3$ simulations are given in Table 5. For the cases with
neighbor size of $n\leq 8$, the substructures are still significant with the
probabilities $P(\kappa_n>\kappa_n^{obs}) < $10\%.  The degree of difference
between the local velocity distribution for groups of six nearest neighbors
and the overall velocity distribution is shown by the bubble plot in Fig. 11b.
Compared with Fig. 5b, substructure A becomes less conspicuous, and
substructures B and C remain with similar significance. Additionally, no
clustering of large bubbles are found at the positions of clumps D and E in
Fig. 11b, indicating that clumps D and E may not be real substructures in
velocity distribution.

Within the areas of clumps A, B and C, defined by the contour curve at 0.15
arcmin$^{-2}$ in Fig. 5a, there are 10, 17 and 8 galaxies, respectively. Table
6 shows the statistics of line-of-sight velocity distributions for these
galaxies. Because of the small number of galaxies in the substructures, we
used the ROSTAT software to compute the biweight locations and scales. It is
interesting to find that subclusters A and C locate at the opposite sides of
the central concentration, and they are likely to have reverse distribution of
peculiar velocities, respective to the systemic redshift of 0.0806. If we
regard the biweight location of the velocity distribution of 214 known member
galaxies ($C_{BI}=24025\pm89$) as systemic receding velocity, subcluster A
tends to move toward us at more than 99\% confidence level, and subcluster C
tends to depart from us at nearly 2$\sigma$ significance. Additionally
substructure B has a slight tendency similar to substructure C. This might be
a direct evidence for an unrelaxed system, and supports a recent merge event.
We applied the same statistics to the spectroscopically confirmed galaxies in
these three substructures, and similar dynamical picture was achieved. This
should be the first report of the direct dynamical evidence in optical band
for an on-going merger in Abell~2255.

%table 3
\begin{table}[ht]
%\caption[]
\noindent {Table 5: Result of $\kappa$-Test for 527 Cluster Galaxies}

\vspace{2mm}
\begin{tabular}{ccccccccc}   \hline   \hline
\noalign{\smallskip}  
 $n$ & 
5      &       6&      7&      8&      9&     12&     15&     20\\
\noalign{\smallskip} \hline \noalign{\smallskip}
$P(\kappa_n>\kappa_n^{obs})$ & 
0.026  &  0.017 &  0.057&  0.097&  0.133&  0.174&  0.184&  0.223\\
\noalign{\smallskip}   \hline  
\end{tabular}
\end{table}

%table 3
\begin{table}[ht]
%\caption[]
\noindent {Table 6: Properties of Substructures A, B, and C}

\vspace{2mm}
\begin{tabular}{llll}   \hline   \hline
\noalign{\smallskip}
 Statistic         & Substructure A & Substructure B & Substructure C \\
 \noalign{\smallskip} \hline \noalign{\smallskip}
 Size (known+new)     &  6+4           &  10+7             &  7+1             \\
 Galaxy No. in Table 2 & 116,123,126,133,   & 118,129,130,137,140,&
138,154,166,169\\
                 & 134,148            & 142,152,153,162,167 &172,173,178 \\
 Galaxy No. in Table 4 & 266,268,269,271    & 41,42,44,61,62,63,72 & 27
\\ Mean Velocity (\kms) & 23190 $\pm$ 528 & 24408 $\pm$ 448 & 24711 $\pm$ 415\\
Standard Deviation (\kms)&  1668      &    1848          &           1173  \\
Biweight Location (\kms) & 23109 $\pm$ 257& 24400 $\pm$ 482 & 24893 $\pm$ 459
\\ Biweight Scale (\kms)     &  748 $\pm$ 599 & 1908 $\pm$ 220 & 1144 $\pm$ 542
  \\
\noalign{\smallskip}   \hline
\end{tabular}
\end{table}

\subsection{\sl Color Properties and Luminosity Function}

The C-M relations for the enlarged sample are shown in Fig. 12, spanning a
wider range of the $h$ magnitude from 15.0 to 20.5. The SDSS spectroscopic
target selection of the main galaxy sample is complete to $r'<17.77$ (Strauss
et al. 2002), which has been well shown in Fig. 12. The linear relationship
defined by the bright known early-type galaxies can well be extended to the
faint early-type galaxies. This correlation can be understood by the fact that
the reddening of galaxies is mainly caused by the rich metal absorption in
stellar spectra shortward of 4250 \AA, and it tends to be more significant for
bright early-type galaxies. Such a correlation was also found for the
early-type galaxies in Virgo and Coma clusters, and elliptical galaxies seem
to have a tighter correlation than the lenticular galaxies do (Bower et al.
1992). Furthermore, the late-type galaxies have a larger dispersion of color
indices than the early-types (Yuan, et al.2001).

\vskip 5mm \centerline{\framebox[12cm][c]{Fig. 12: The C-M relations for all
cluster galaxies}}

The luminosity function (LF) in a cluster is a powerful and fundamental means
to study galaxy formation and evolution in a dense environment. Recent studies
showed that the LF for cluster galaxies differ from one cluster to another,
and even from one region to another region in one cluster (Biviano et al.
1995; Durret et al. 2002). Our photometry for all member galaxies in
Abell~2255 provide a good sample for studying the LF for the bright galaxies
in a rich cluster. Some previous studies on nearby rich clusters show that
there may be at least two components in the galaxy LF, a Gaussian distribution
for the bright galaxies and another component for fainter cluster galaxies
formulated by the Schechter function or a power law (Durret et al. 1999). For
comparison with other investigations, we have chosen to plot the LF in
traditional $R$- and $V$-band magnitudes which can be calculated by
$R=i+0.1036 (\pm 0.055)$ and $V = g + 0.3292(f-h) + 0.0476(\pm 0.027)$ (Zhou,
et al. 2003). Fig. 13 shows the distribution of $R$ and $V$ magnitudes for all
galaxies in Abell~2255. The LFs are significantly asymmetric, with a peak at
$R \sim 17.5$ and $V \sim 17.7 $ respectively. The vast majority of bright
member galaxies ($R<17.5$ or $V<17.8$) has been confirmed by the SDSS
spectroscopies. It is clear that these previously known data-set may not
strongly constrain the LF shape because the peak of LF can not be well
determined.

The application of the photometric redshift technique to 254 known galaxies
shows that the selection criterion $0.068<z_{phot}<0.090$ excludes nearly 10\%
of the true members, and leads to about 5\% contamination by background
galaxies. The success rate for membership selection based on our SED-fitting
procedures should be still over 80\% for the bright galaxies. Though the
uncertainty in membership selection might be larger for faint galaxies, the LF
peak shown in Fig. 13 should be real because the BATC photometry are complete
for the galaxies brighter than 19.0 mag in all filter bands (see Table 1). The
magnitude distribution for the cluster galaxies brighter than 19.0 mag can be
expected to put a strong constraint on the true LF shape. Therefore we used
the minimum $\chi^2$ method to fit the bright part of the differential LF with
(i) a Schechter function: $S(m)= K_S\, 10^{0.4(m^*-m) (\alpha+1)}\,
exp[-10^{0.4(m^*-m)}]$, and (ii) a Gaussian function: $G(m) = K_G\,
exp[-(m-\mu_m)^2/(2\sigma_m^2)]$. The best fit parameters are also shown in
Fig. 13.

The accuracy of $z_{phot}$ estimate is comparatively high for the brighter
galaxies that have smaller magnitude errors, so the bright member galaxies
have higher possibility to be selected. The bright part of the LF can be
fitted well with a single Gaussian defined by a central absolute magnitude of
$\mu_{M_V}=-20.3(\pm0.1)$ and a dispersion of $\sigma \sim 1.02(\pm0.09)$,
which is consistent with the Gaussian component for Coma LF,
$\mu_{M_V}=-20.4(\pm0.2)$ and $\sigma = 1.1 (\pm0.3)$ (Lobo, et al. 1997). A
better fit to the faint part of the LF can be obtained using a single
Schechter function with $M_V^* = -20.58 (\pm0.25)$ and $\alpha = -0.136
(\pm0.275)$, though there exists an excess in the bright end of LFs. The LF
curve for the cluster galaxies fainter than 19.0 mag has a significant excess
above the Gaussian fitting, and another component of a Schechter function
seems to be need to give a better fit to the faint part. However, our sample
for dwarf cluster galaxies is far from complete because a large number of
faint galaxies can not be detected by the BATC multicolor system. The lack of
BATC SED information prevents an accurate membership selection for dwarf
galaxies in Abell~2255.

%A conservative estimate for the completeness of the cluster galaxies should
%be over 60\% at $R = 19.0$.

\vskip 5mm \centerline{\framebox[12cm][c]{Fig. 13: Distributions of $R-$ and
$V-$ magnitudes}}

\section{SUMMARY}

This paper presents our multicolor optical photometry for the nearby rich
cluster of galaxies Abell~2255, using the 60/90 cm Schmidt Telescope of the
BAO equipped with 13 BATC filters that cover almost whole optical wavelength
domain. A method of SED combination for the SDSS and BATC photometries has
been explored. After an aperture correction of the SDSS {\em model} magnitudes
and a flux calibration of the BATC magnitudes, the SEDs obtained by these two
photometric systems were carefully combined. A sample of 254 galaxies with
known spectroscopic redshifts in the region of Abell~2255 was then formed for
verifying the reliability of the combined SEDs. We applied the technique of
photometric redshift upon this sample, and more than 90\% of the member
galaxies are found to have photometric redshifts between 0.068 and 0.090,
which provides a reliable tool for further membership selection. This results
showed that the combined SEDs with higher resolution could lead to a more
accurate estimate of photometric redshift. We selected 214 known member
galaxies from our SED lists, for which the detailed analyses on their spatial
and color properties are performed. We found that the core region of
Abell~2255 is populated by early-type galaxies, and the late-type galaxies are
scattered. 137 known early-type galaxies are found to have a tight
color-magnitude correlation.

The cross-identification of the SDSS-listed extended sources with our SED list
yields a large sample which contains 2522 faint galaxies without redshift
information. Based on the knowledge of SED features for known cluster
galaxies, we performed our SED-fitting procedures for estimating the
photometric redshifts for these faint galaxies. The color-magnitude
correlation was also used to constrain our membership selection. As a result,
we isolated 313 galaxies as the faint members. The C-M relation for this
enlarged sample of cluster galaxies shows that the tight correlation defined
by bright galaxies can be extended to the faint early-type galaxies, and the
late-type galaxies have a larger dispersion of color index. Our sample of 527
cluster galaxies is nearly complete up to $V=19.0$, one magnitude fainter than
the LF maximum, which could put a strong constraint on the LF shape.
Subclusters A and C located on opposite sides of the cluster center are found
to have reverse tendencies in their peculiar velocities, which obviously
supports the picture of cluster/group merger in this cluster, as proposed by
the radio and X-ray observations.

It should be noted that the combined SED measures through a fixed aperture
offers a higher spectral resolution for all the objects detected in the BATC
and SDSS surveys, which will surely benefit each other. Our candidate list of
faint member galaxies provides the targets for follow-up SDSS spectroscopic
observations. For a further investigation on dynamics and luminosity function
of galaxies in such a rich cluster possessing an on-going cluster/group
merger, the deep spectroscopic observations for the sake of obtaining the
accurate redshifts for the faint member galaxies is necessary.

%\vskip 1cm

\acknowledgments

We acknowledge the anonymous referee for his thorough reading of this paper
and invaluable suggestions. We are grateful to Dr. Beers, R. C. and Ashman, K.
M. for their kindly providing the ROSTAT software and the KMM algorithm code
which has been used in this work. We have made use of the NASA/IPAC
Extragalactic Database (NED) which is operated by the Jet Propulsion
Laboratory, California Institute of Technology, under contract with the
National Aeronautics and Space Administration. This work is mainly supported
by the National Key Base Sciences Research Foundation under the contract
TG1999075402 and is also supported by the Chinese National Science Foundation
(NSF) under the contract No.10273007. We would like to thank Prof. Jiansheng
Chen, Dr. Jun Ma, Hong Wu and Yanbin Yang for their helpful discussions, and
Prof. Helmut Abt for his help in improving the English.  We also appreciate
the assistants who contributed their hard work to the observations.

%\clearpage

%% -----------------------------------------------------------

\clearpage

%fig1
\begin{figure}
\epsscale{0.65} \plotone{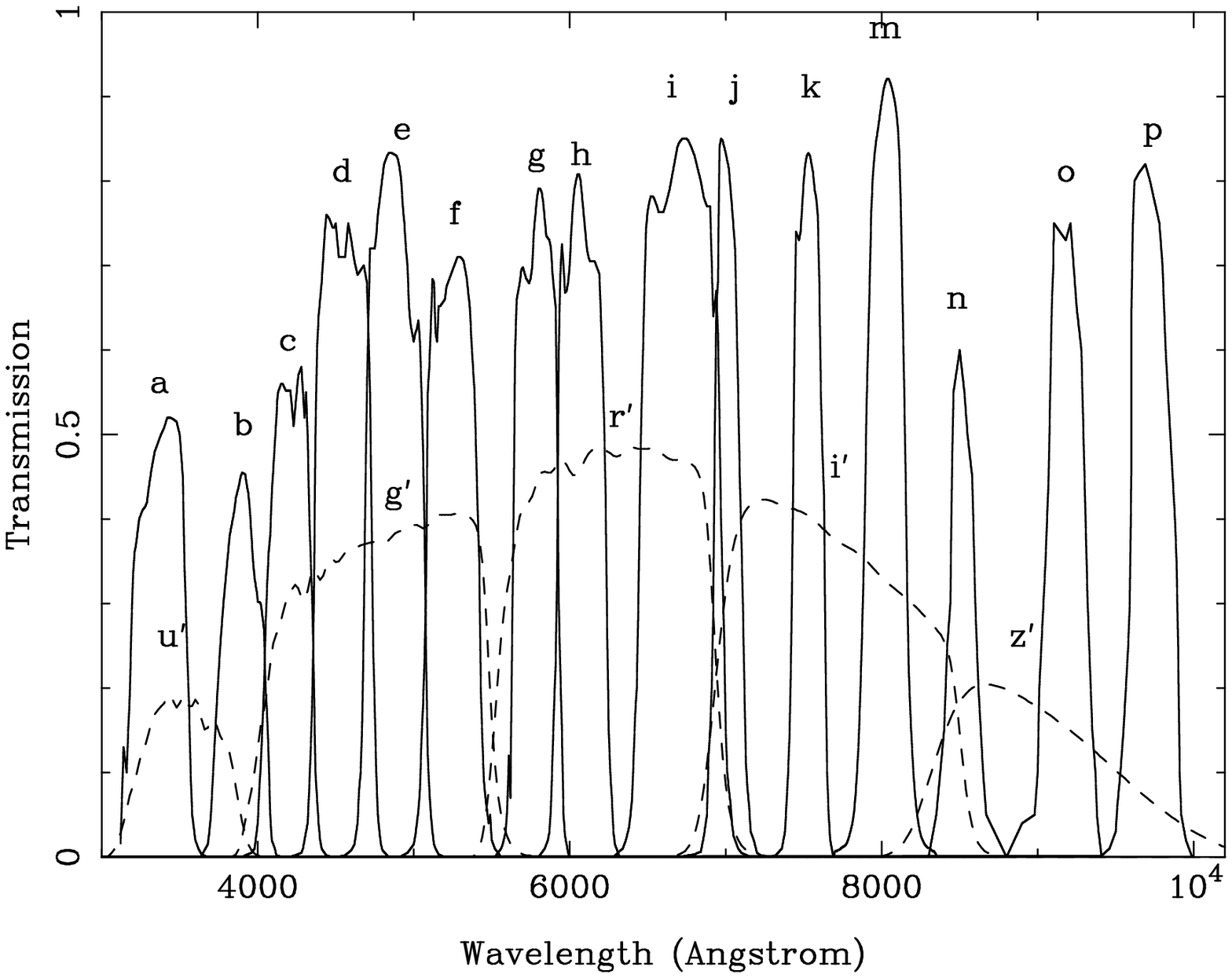} \caption{The transmission curves of the BATC
and SDSS filters. The name of each filter is labelled on the top of
transmission curve.}
\end{figure}

%\clearpage

%fig2
\begin{figure}
\epsscale{0.65} \plotone{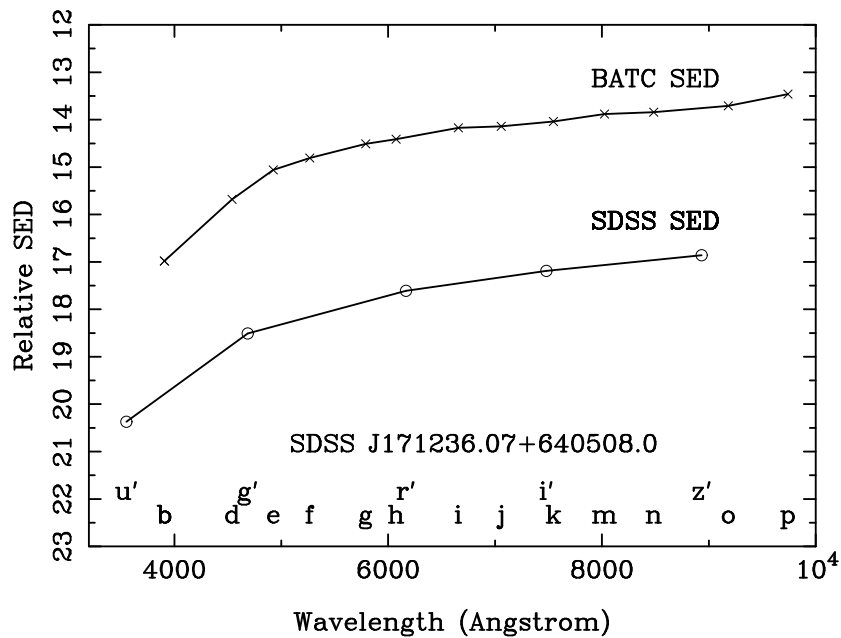} \caption{The relative SEDs in the BATC and
SDSS multicolor systems for the source SDSS J171236.07+640508.0, the central
early-type galaxy in Abell~2255 cluster.}
\end{figure}

%\vskip 1cm
%\clearpage

%fig3
\begin{figure}
\epsscale{0.65} \plotone{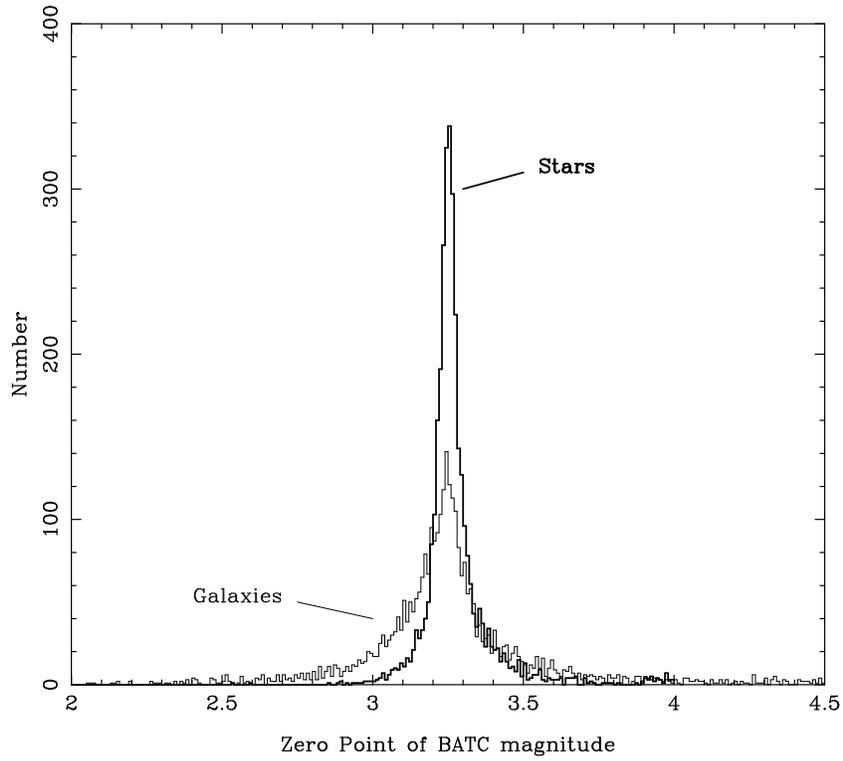} \caption{Distribution of the zero-points for
the stars and galaxies detected by both multicolor surveys.}
\end{figure}

%fig4
\begin{figure}
\epsscale{1.0} \plotone{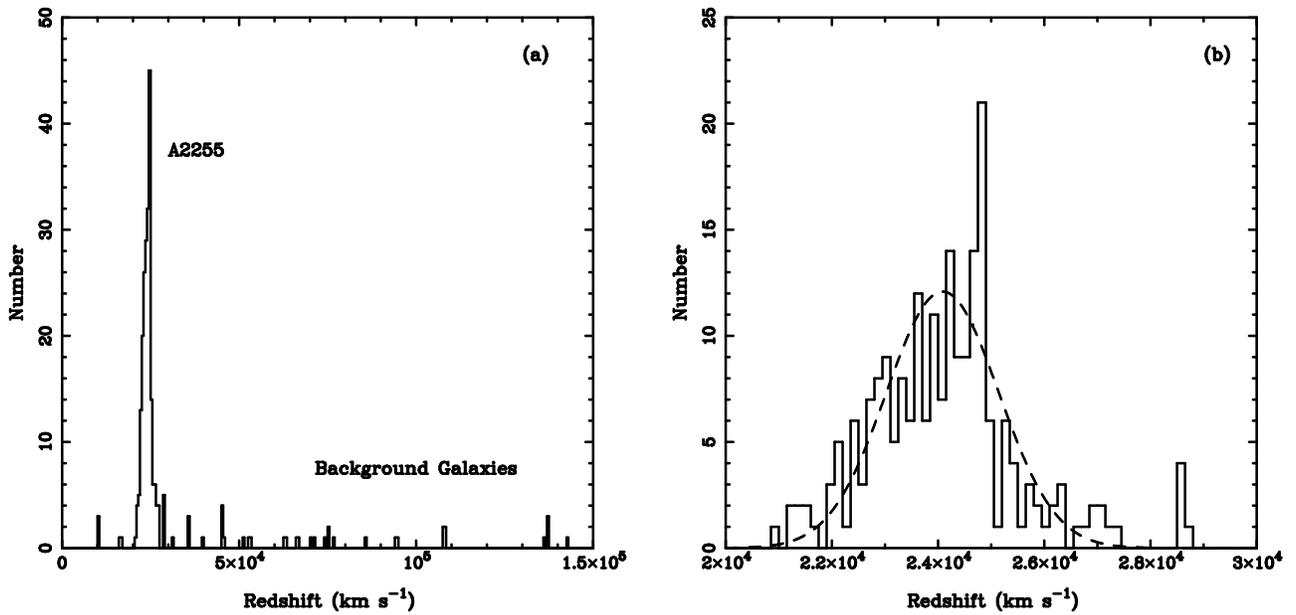} \caption{ Distribution of spectroscopic
redshifts for (a) 254 known galaxies and (b) 214 member galaxies. The bin
widths are 500 and 150 \kms, respectively. The best-fitting Gaussian model is
also shown in panel (b).}
\end{figure}

%fig5
\begin{figure}
\epsscale{1.0} \plottwo{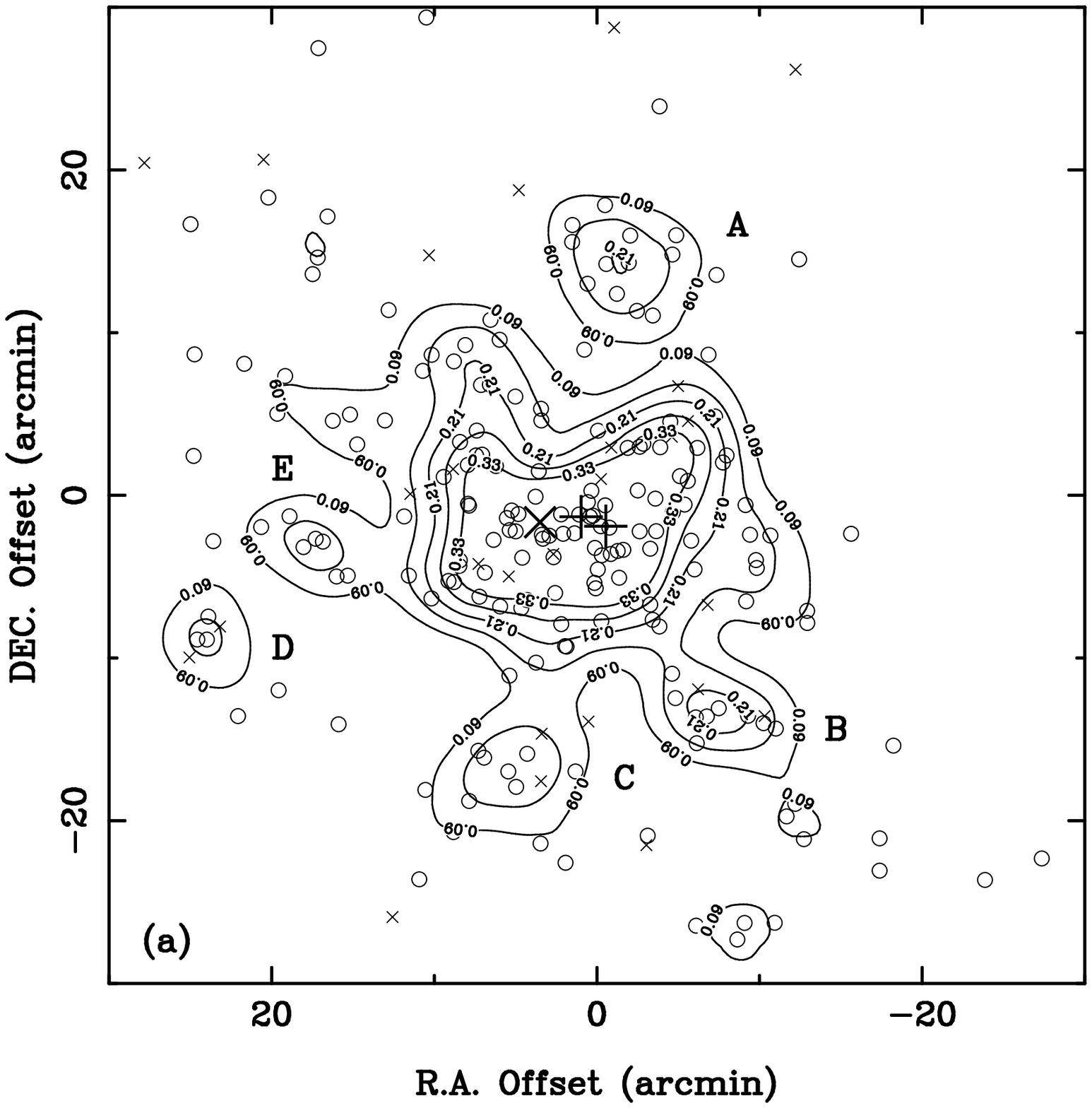}{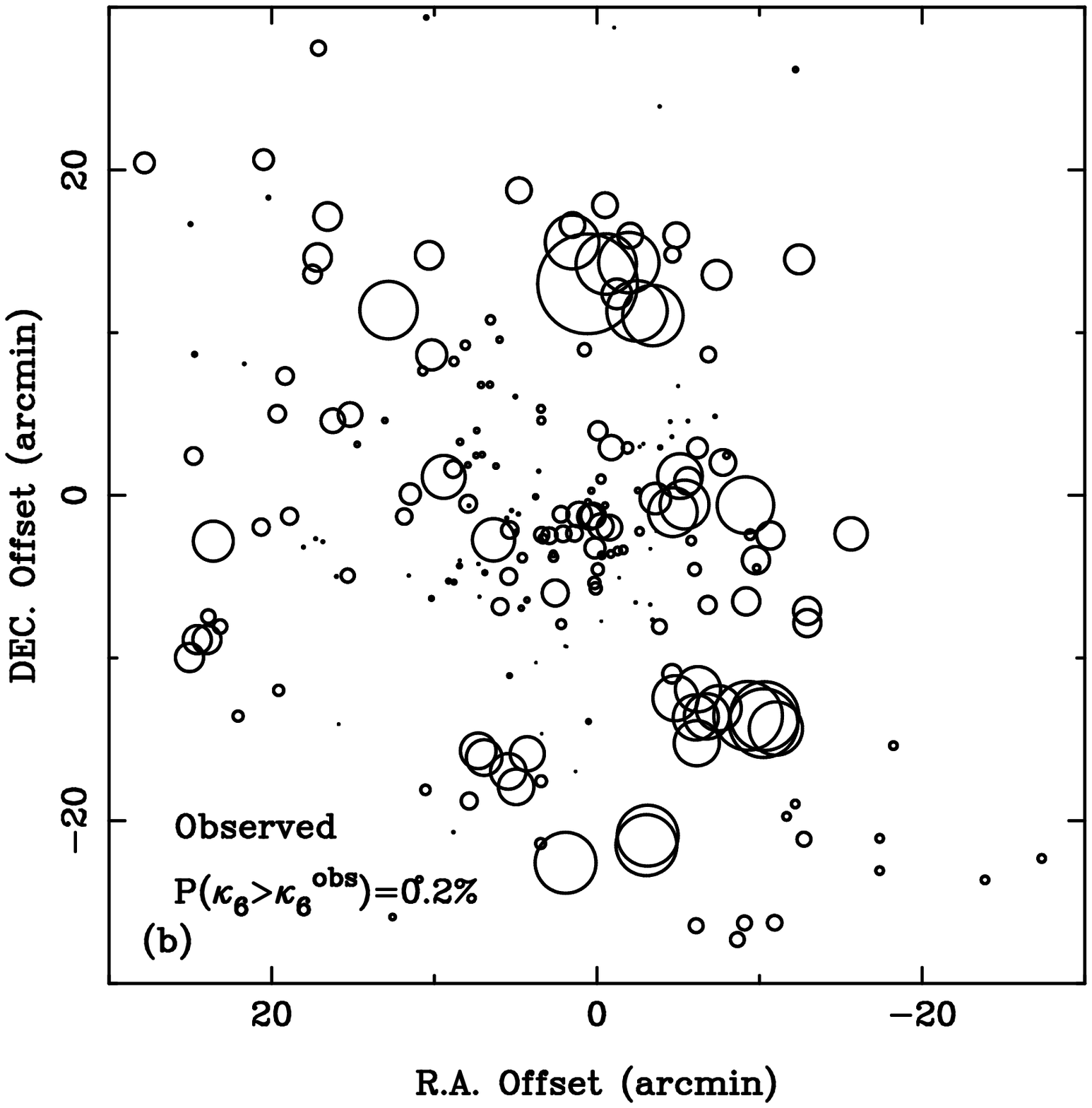} \caption{\small (a) Spatial
distribution of 214 member galaxies, including 168 early-type galaxies ({\em
open circles}) and 46 late-type galaxies ({\em crosses}). The contour map of
the surface density for all these galaxies using the  smoothing window with a
radius of 2 arcmin is also given. The contour levels are 0.09, 0.15, 0.21,
0.27 and 0.33 arcmin$^{-2}$, respectively. The positions for two central cD
galaxies, ZwCl 1710.4+6401 A and B, are flagged as ``$+$'', and the center of
X-ray emission is flagged as ``$\times$''. The contour curves at 0.15
galaxies arcmin$^{-2}$ are used to define the clump areas. (b) Bubble plot
showing the localized variation for groups of six nearest neighbors.}
\end{figure}

%fig.6
\begin{figure}
\epsscale{1.0} \plotone{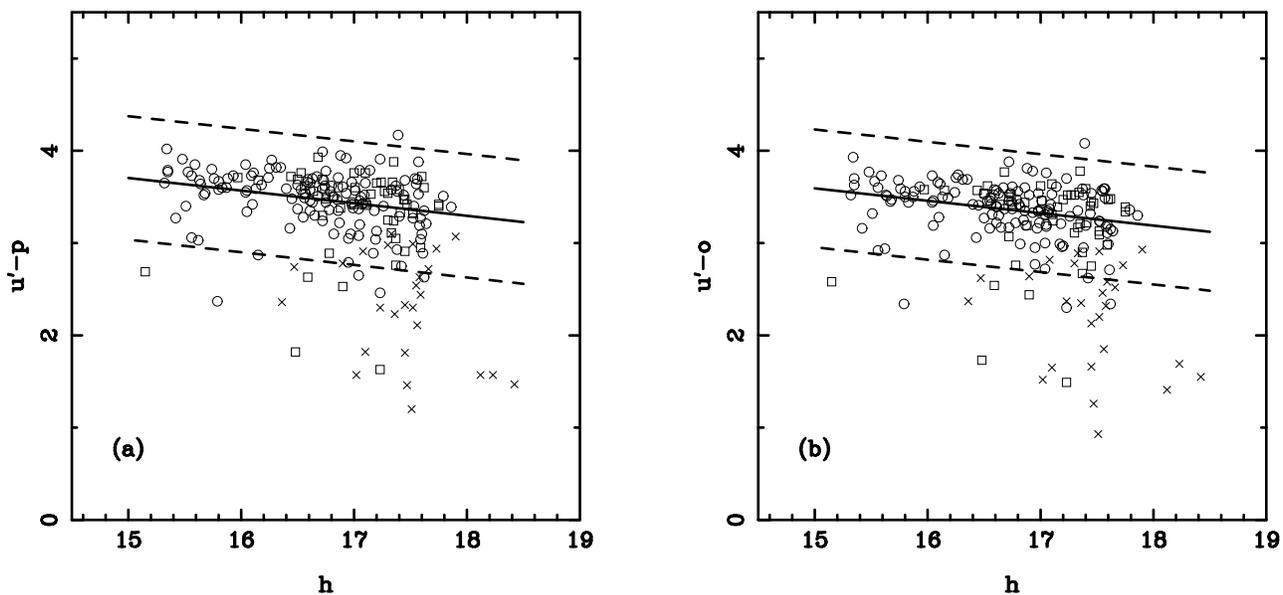} \caption{Color-magnitude relation for known
galaxies in Abell~2255, i.e., the plot of color indices $u'-p$ and $u'-o$
versus $h$ magnitude for 137 elliptical galaxies ({\em open circles}), 51
lenticular galaxies ({\em open squares}), and 26 spiral galaxies ({\em
crosses}).} \end{figure}

%fig.7
\begin{figure}
\epsscale{0.6} \plotone{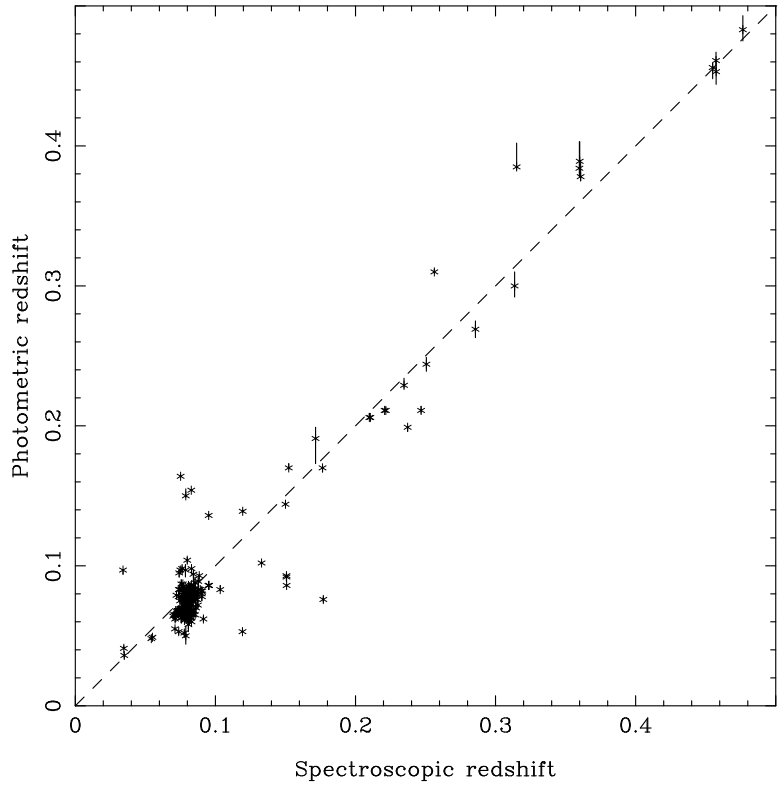} \caption{Comparison between the photometric
redshift ($z_{phot}$) and spectroscopic redshift ($z_{sp}$) for 254 galaxies
with known spectroscopic redshifts in the region of Abell~2255.}
\end{figure}

%fig.8
\begin{figure}
\epsscale{0.6} \plotone{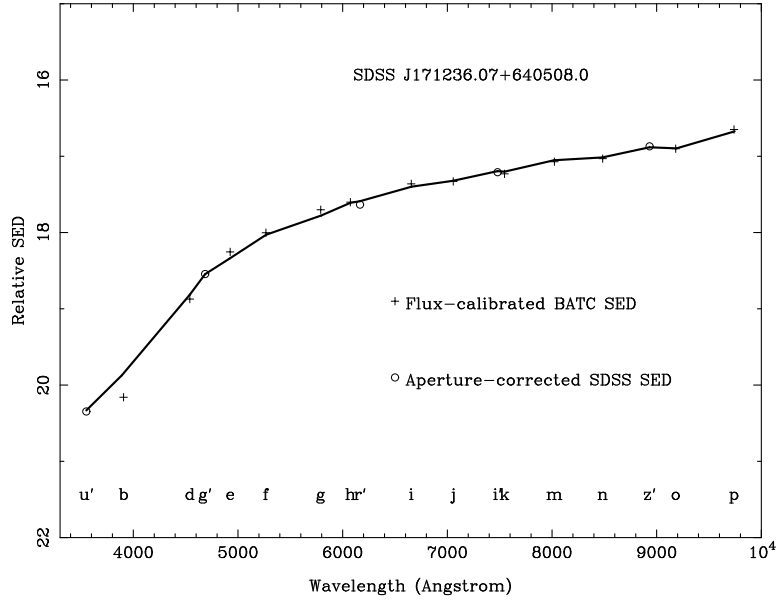} \caption{The best fit of the combined SED of
the source SDSS J171236.07+640508.0, the central early-type galaxy in
Abell~2255 cluster}
\end{figure}

%fig.9
\begin{figure}
\epsscale{1.0} \plotone{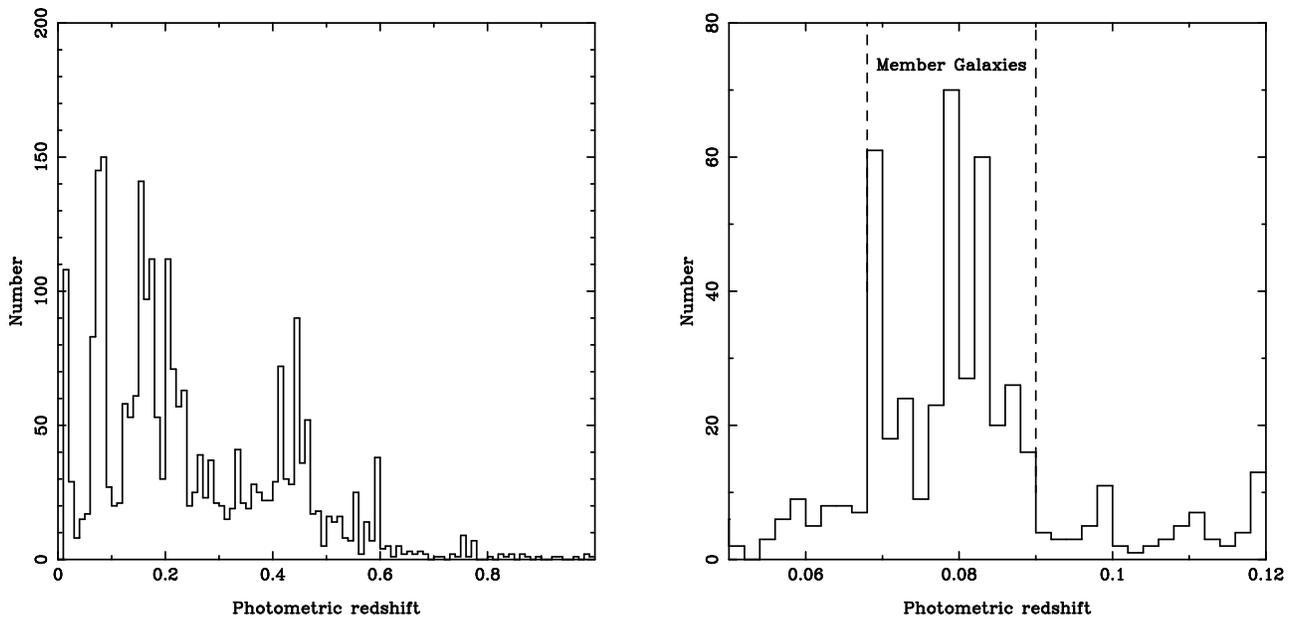} \caption{Distribution of estimated
photometric redshifts for 2522 faint galaxies}
\end{figure}

%fig.10
\begin{figure}
\epsscale{0.5} \plotone{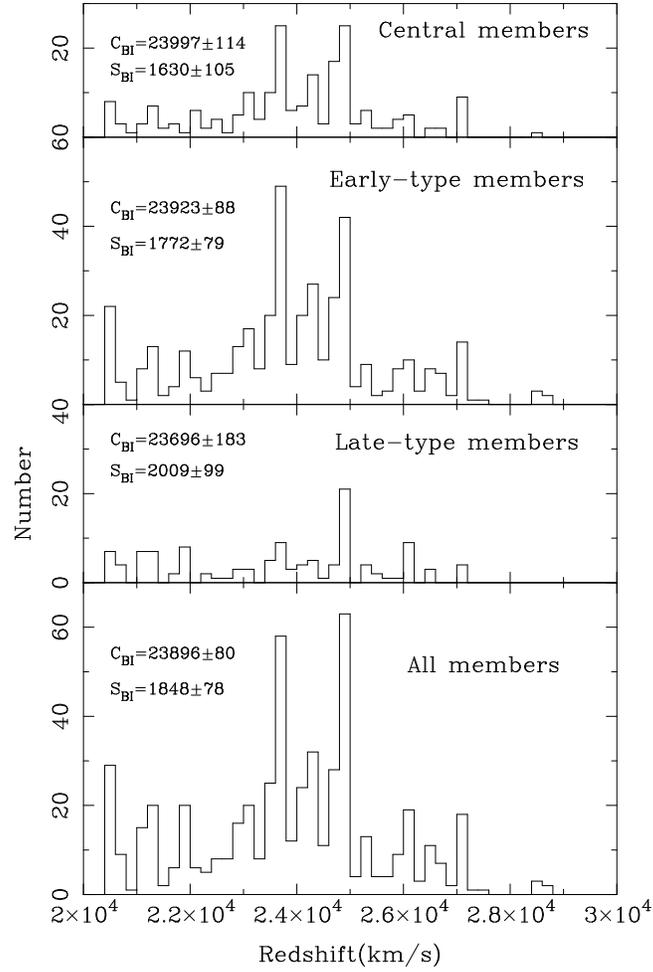} \caption{Redshift distributions for (i) 203
galaxies in central region with a radius of 10 arcmin, (ii) 406 early-type
galaxies, (iii) 121 late-type galaxies, and (iv) the total sample of 527
cluster galaxies.}
\end{figure}

%fig.11
\begin{figure}
\epsscale{1.0} \plottwo{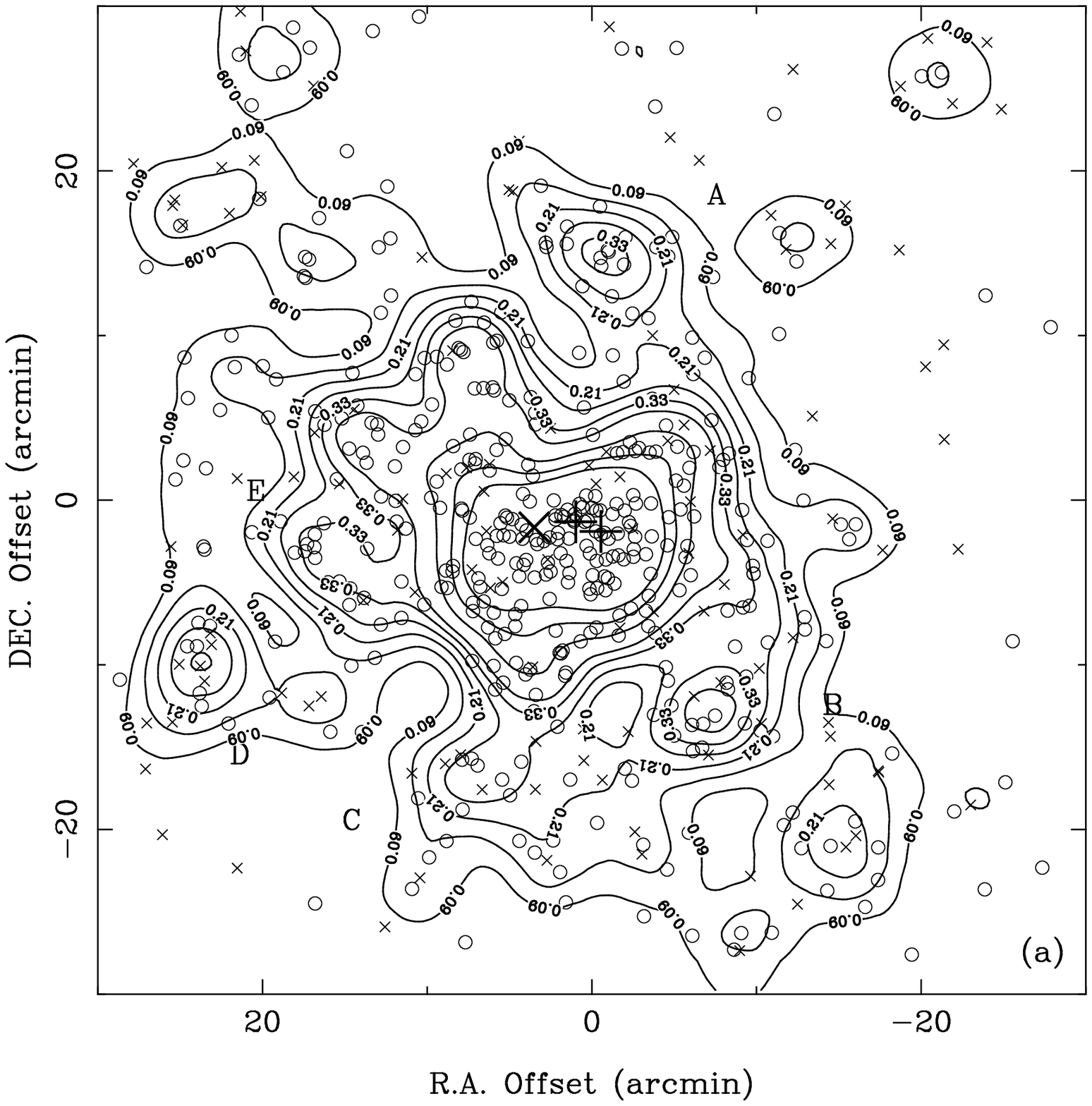}{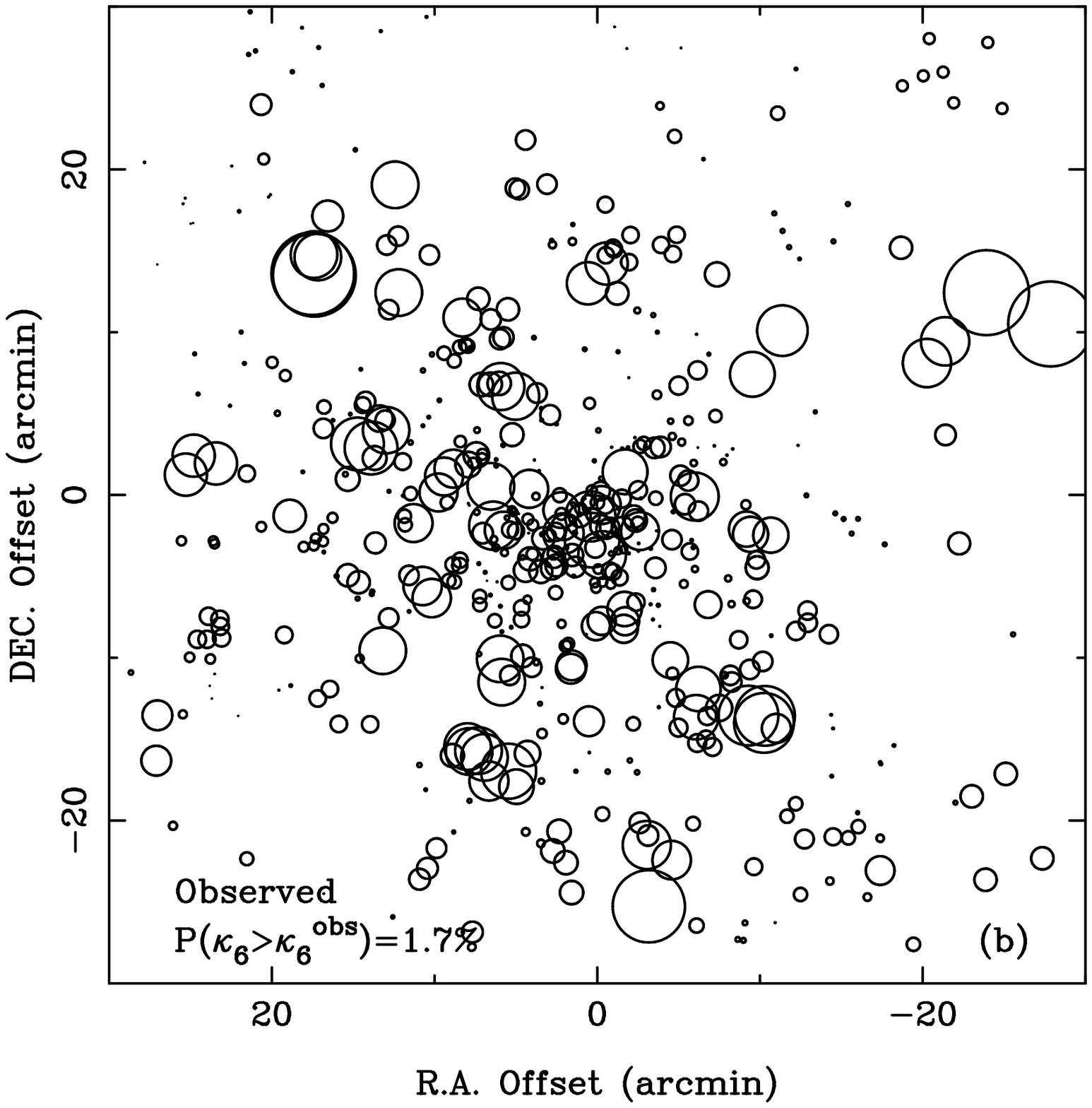} \caption{(a) Spatial
distributions for all member galaxies. Early-type galaxies are denoted by
open circles, and the late-type galaxies by crosses. The contour maps of the
surface density are also superposed. The contour levels are 0.09, 0.15, 0.21,
0.27, 0.33, 0.39, 0.51, 0.63, and 0.75 arcmin$^{-2}$. The positions for two
central cD galaxies, ZwCl 1710.4+6401 A and B, are flagged as `+', and the
center of X-ray emission is flagged as `$\times$'. (b) Bubble plot showing
the degree of difference between the local velocity distribution for groups
of six nearest neighbors and the overall distribution. } \end{figure}

%fig.12
\begin{figure}
\epsscale{0.93} \plotone{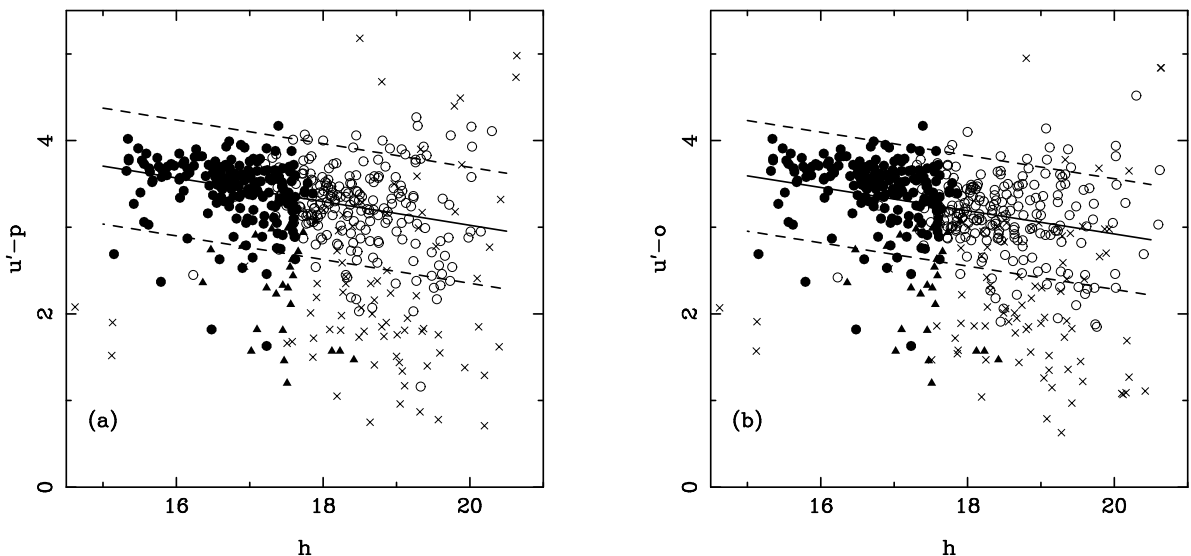} \caption{Color-magnitude relation for all
member galaxies in Abell~2255, i.e., the plots of color indices $u'-p$ and
$u'-o$ versus $h$ magnitude for 188 known early-types (filled circles), 26
known late-types (filled triangles), 188 newly-selected early-types (open
circles) and 46 newly-selected late-types (crosses).} \end{figure}

%fig.13
\begin{figure}
\epsscale{0.95} \plottwo{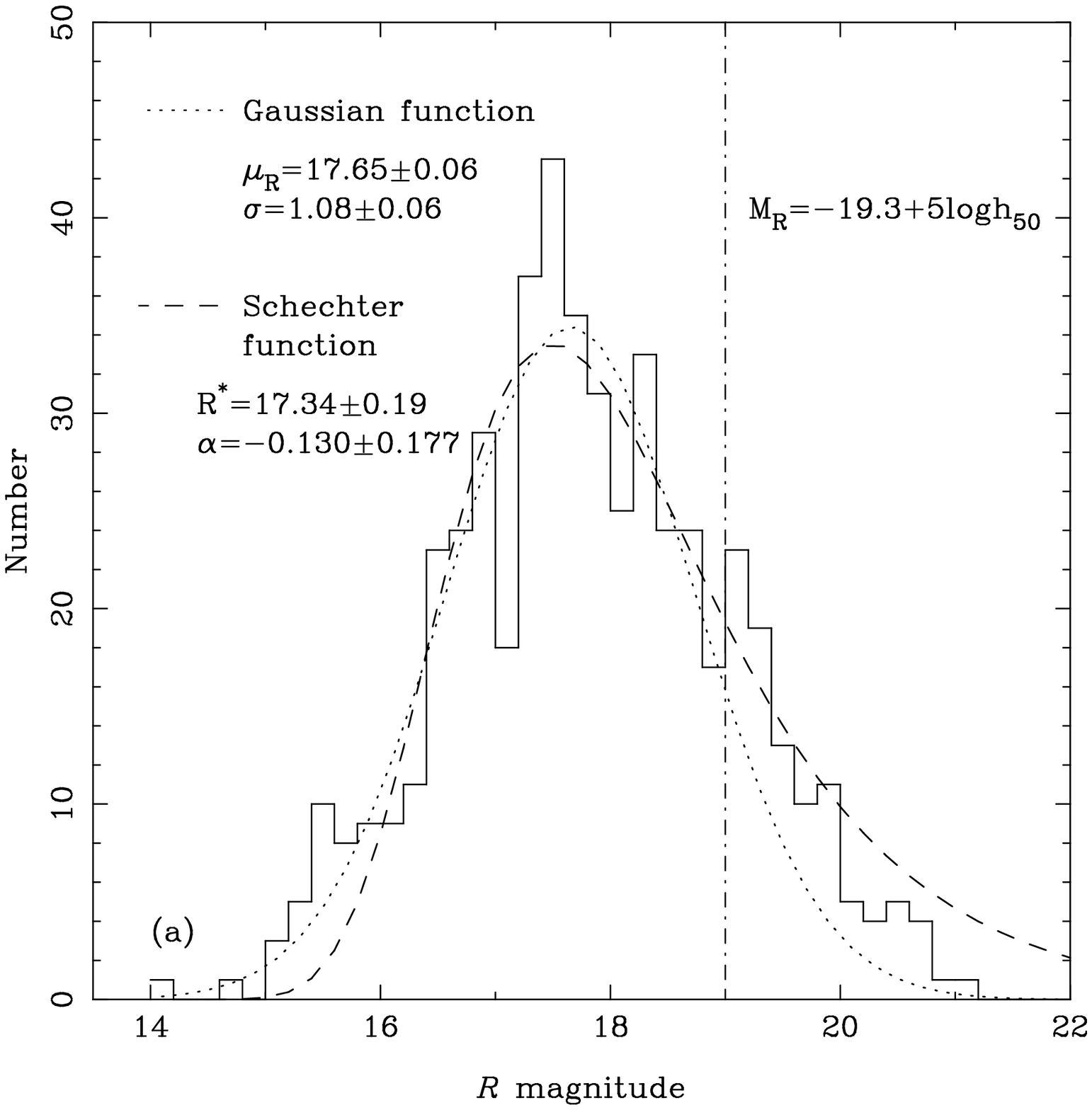}{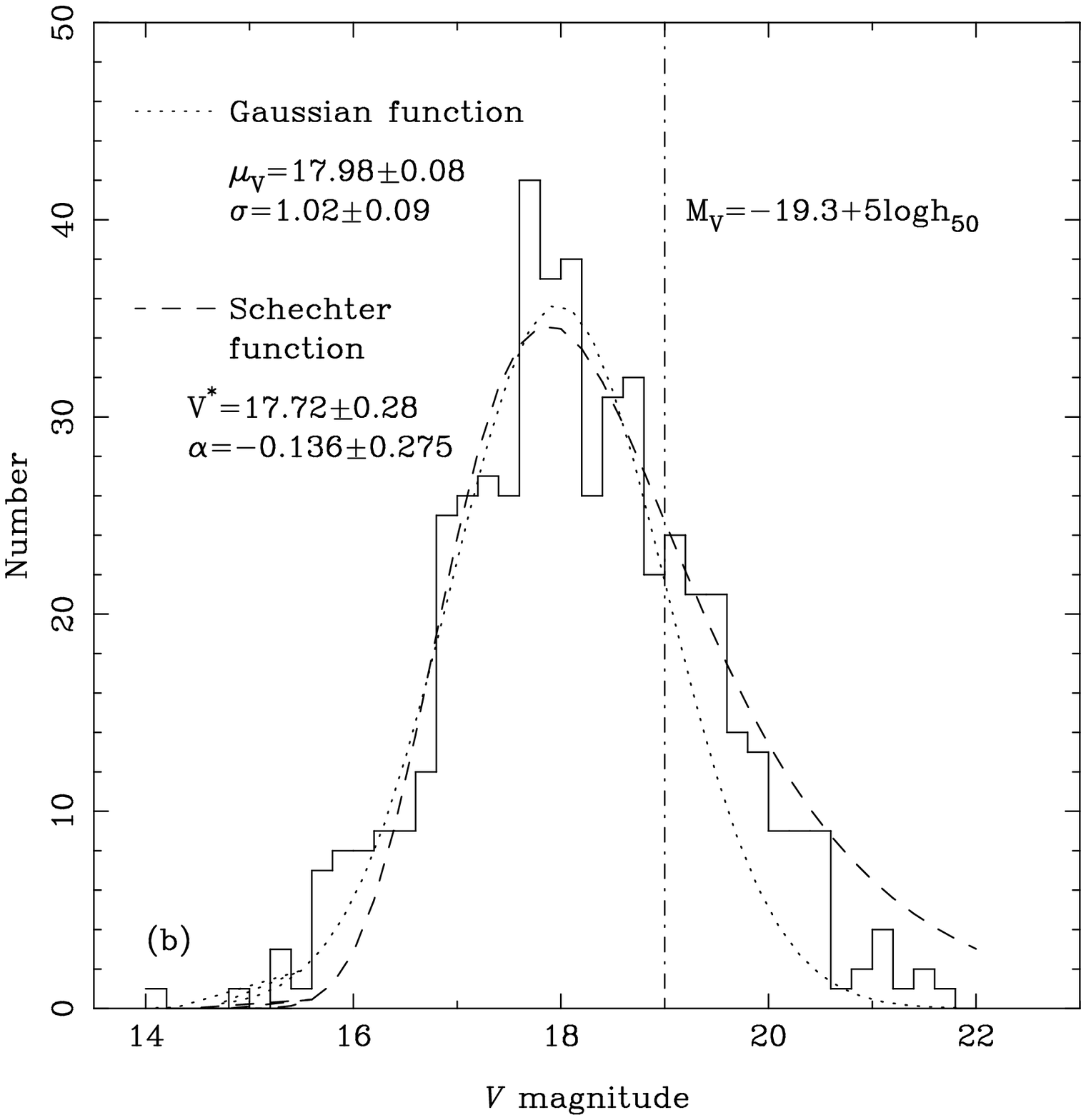} \caption{Distribution of $R$ and
$V$ magnitudes for all cluster galaxies in Abell~2255.}
\end{figure}

%% If you are not including electronic art with your submission, you may
%% mark up your captions using the \figcaption command. See the
%% No more than seven \figcaption commands are allowed per page,
%% so if you have more than seven captions, insert a \clearpage

\clearpage
\def\baselinestretch{1.2}

\begin{figure} \epsscale{1.0}
\plotone{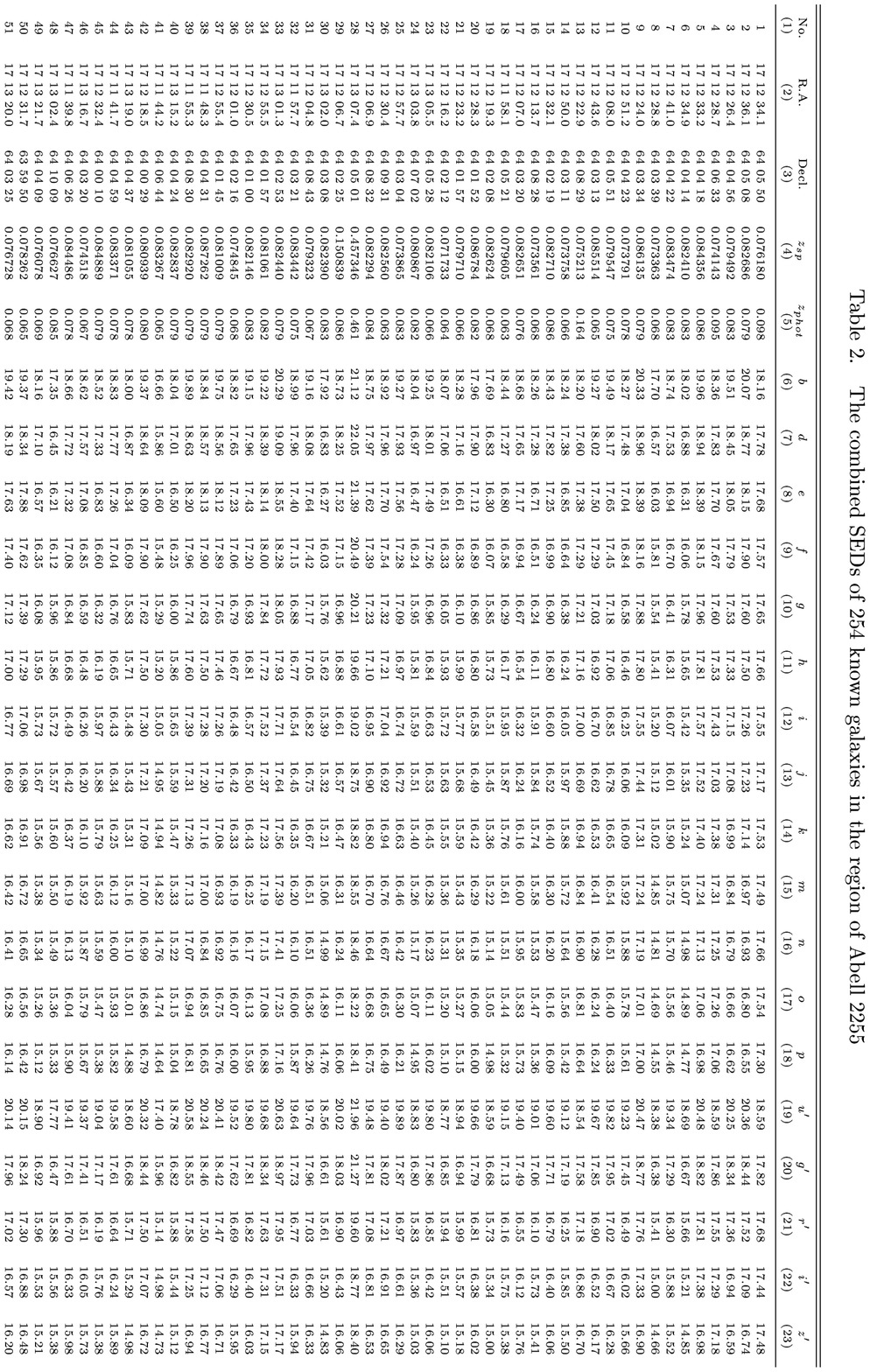} \clearpage
%\caption{}
\end{figure}
\begin{figure}  \epsscale{1.0}
\plotone{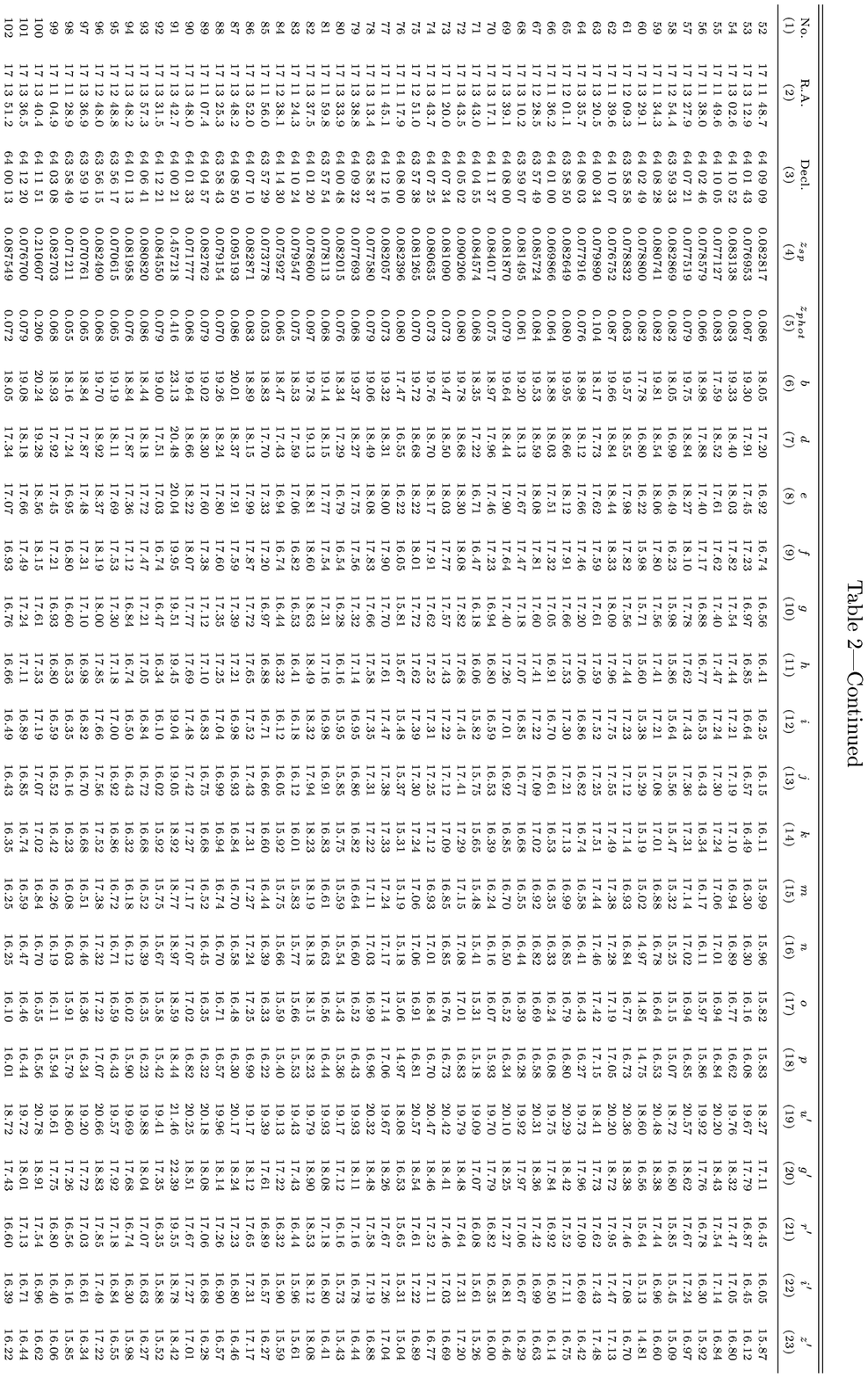}
\end{figure}  \clearpage
\begin{figure} \epsscale{1.0}
\plotone{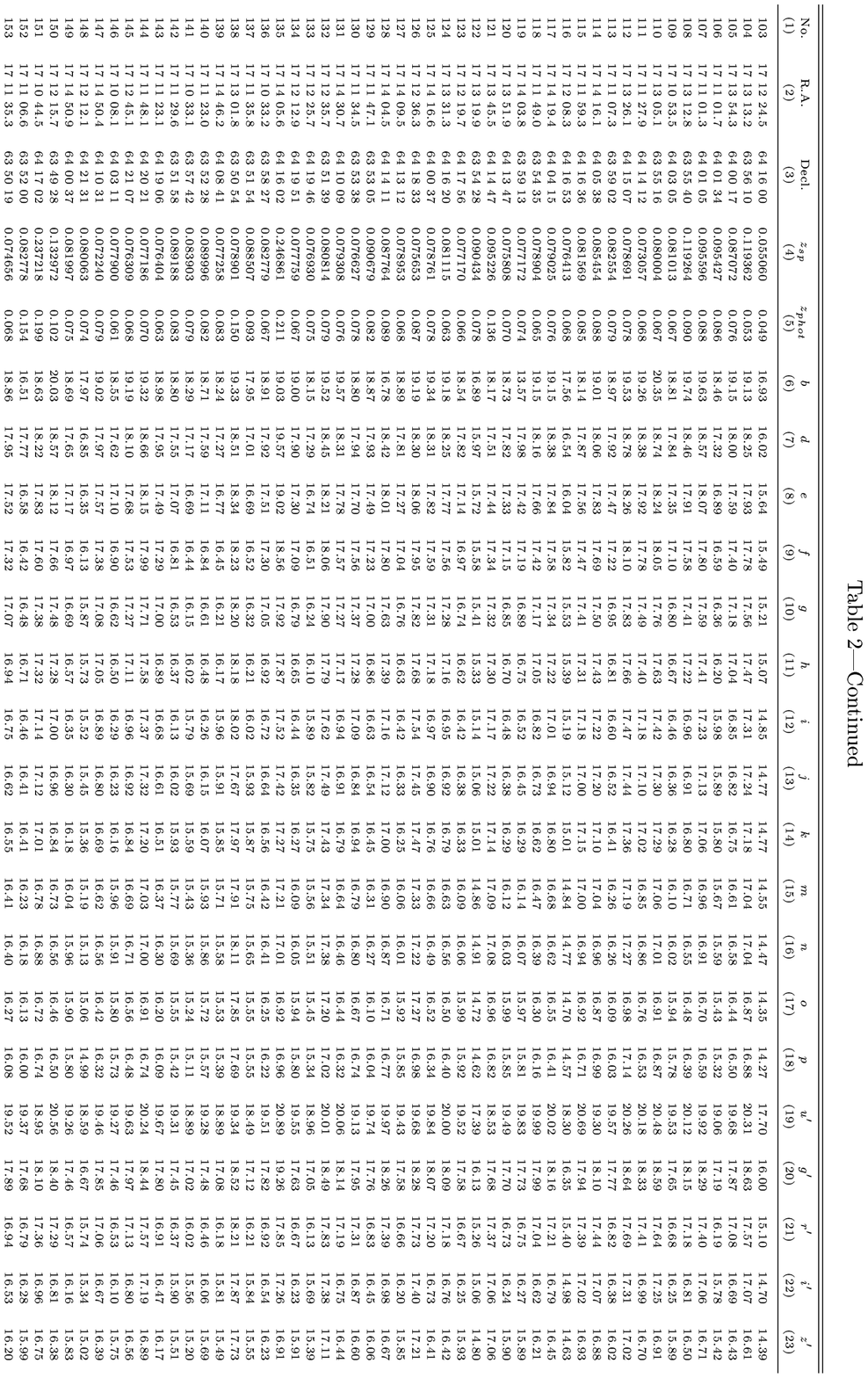}
\end{figure}  \clearpage
\begin{figure}  \epsscale{1.0}
\plotone{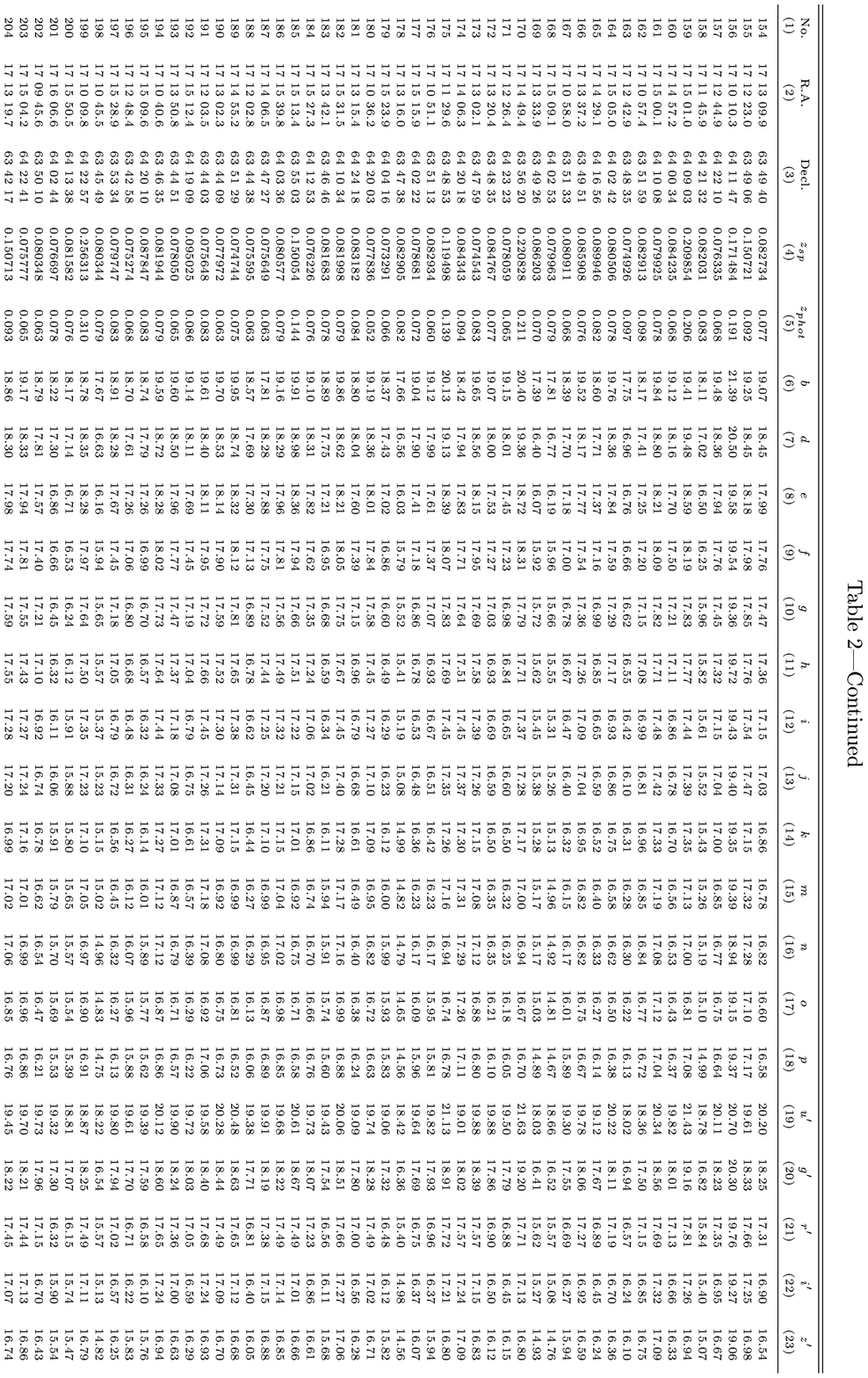}
\end{figure} \clearpage
\begin{figure}   \epsscale{1.0}
\plotone{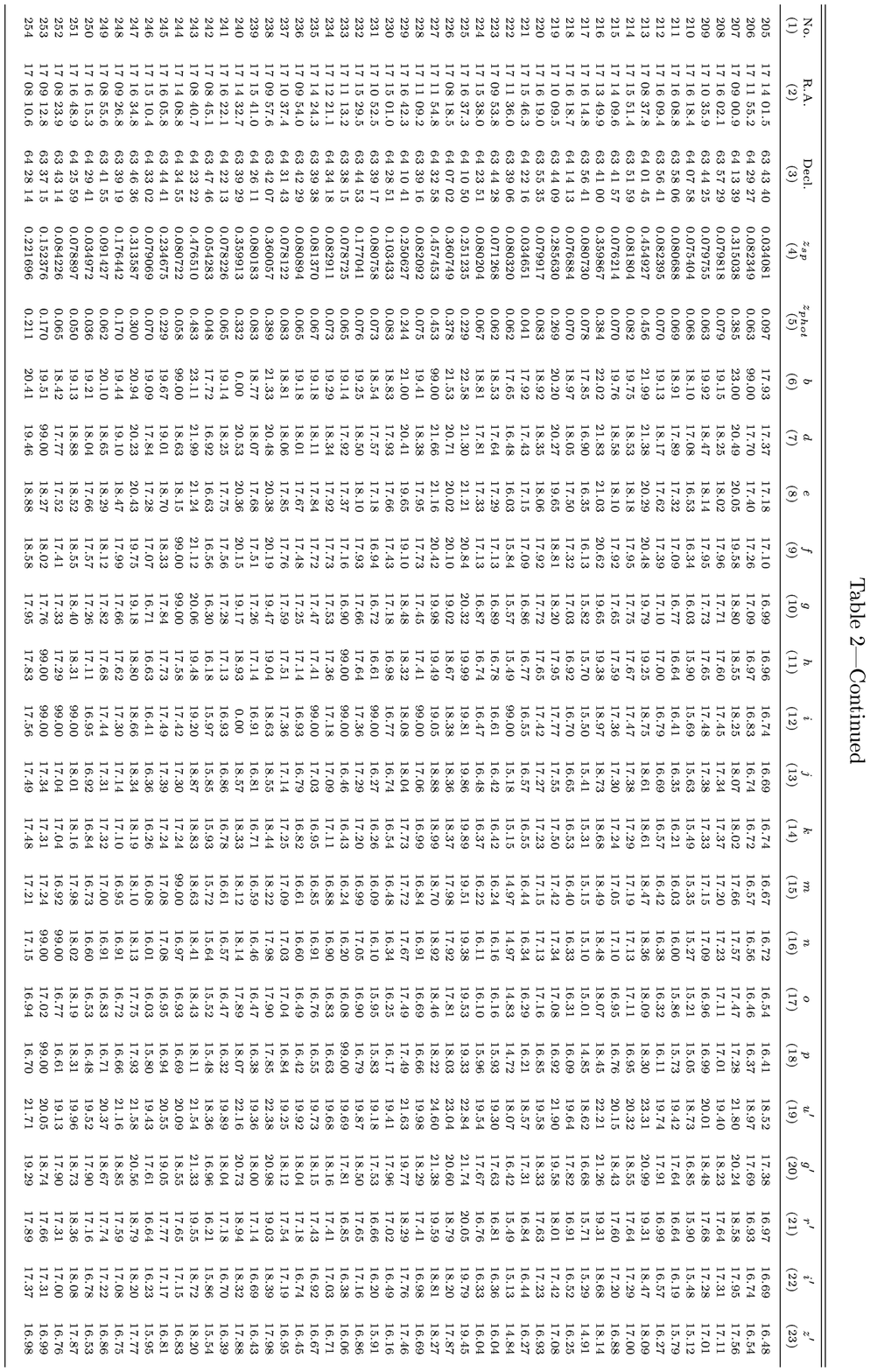}
\end{figure}  \clearpage
%\begin{figure}   \epsscale{1.0}
%\plotone{tab2_6.eps}
%\end{figure}  \clearpage
%\begin{figure}   \epsscale{1.0}
%\plotone{tab2_7.eps}
%\end{figure} \clearpage

%\input{tab3.tex}

\begin{figure}  \epsscale{1.0}
\plotone{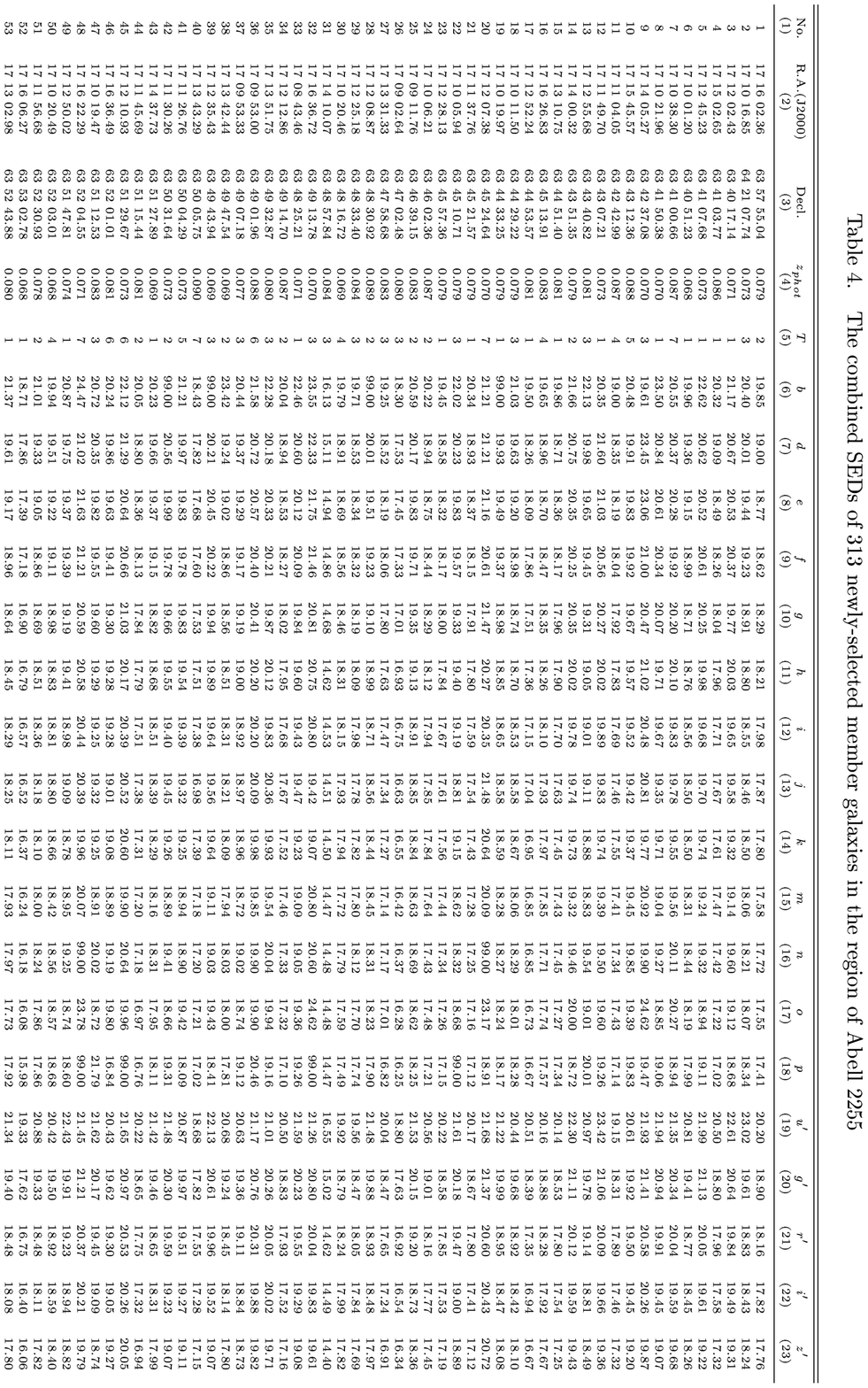}
\end{figure}  \clearpage
\begin{figure} \epsscale{1.0}
\plotone{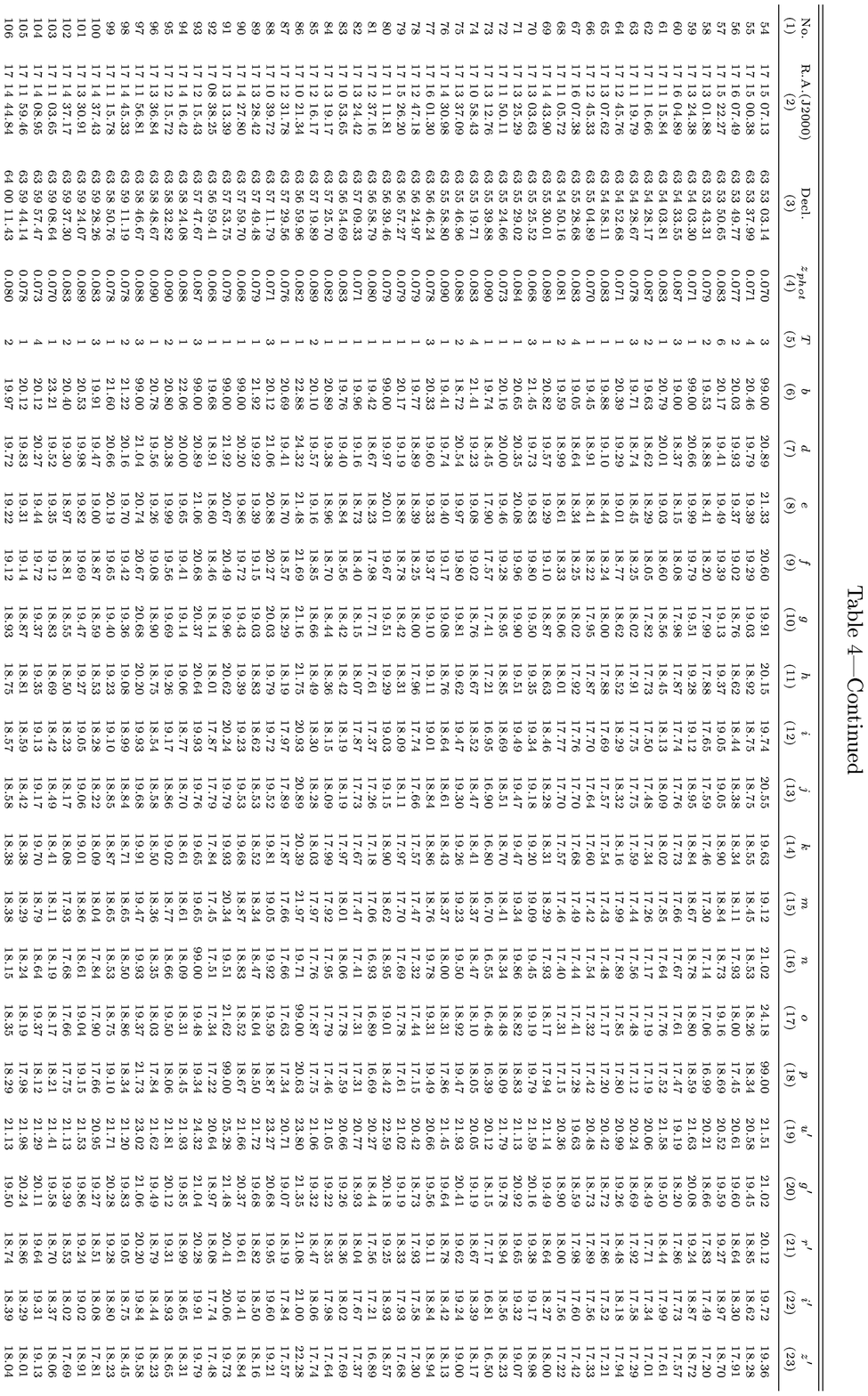}
\end{figure}  \clearpage
\begin{figure}  \epsscale{1.0}
\plotone{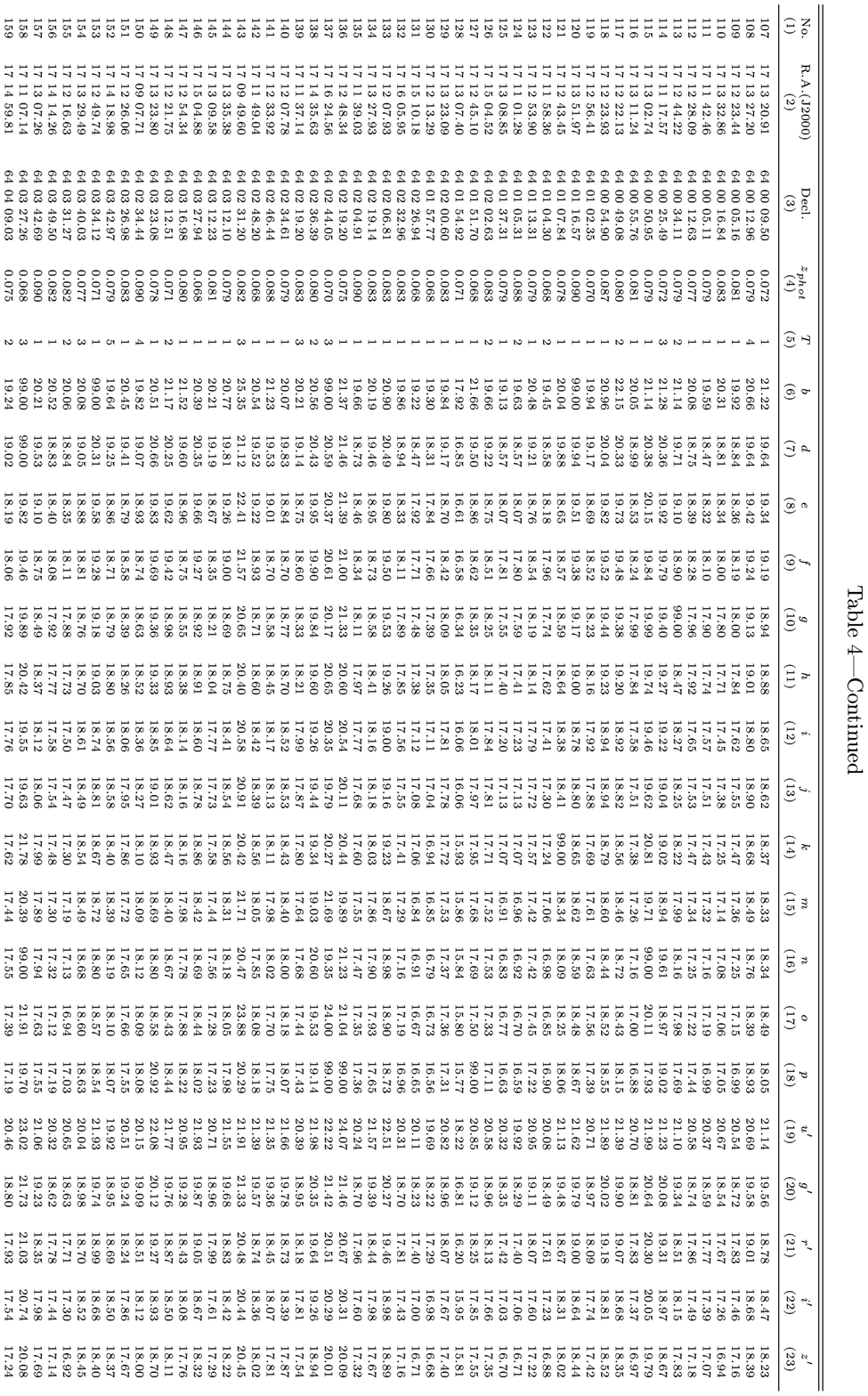}
\end{figure} \clearpage
\begin{figure}   \epsscale{1.0}
\plotone{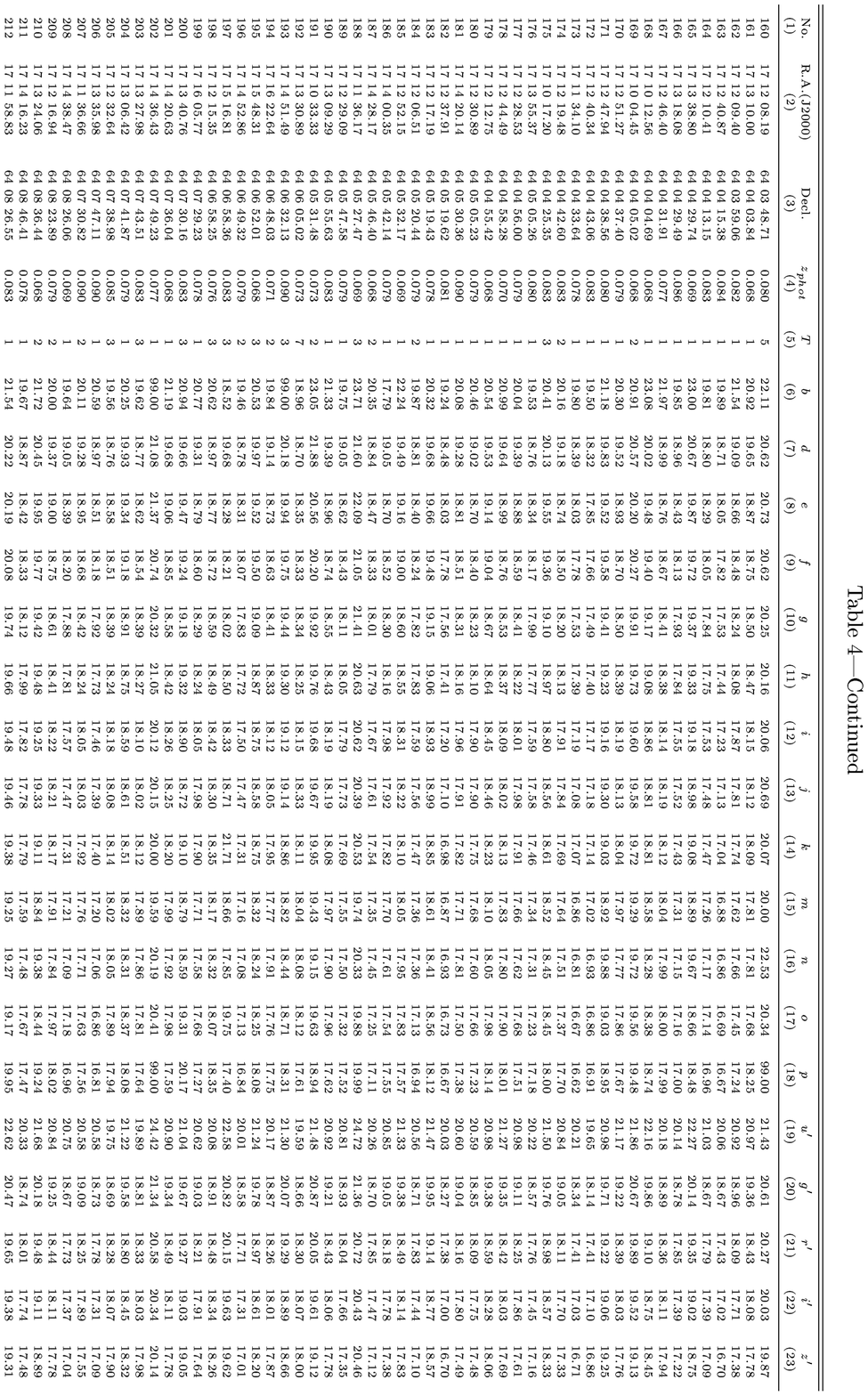}
\end{figure}  \clearpage
\begin{figure}   \epsscale{1.0}
\plotone{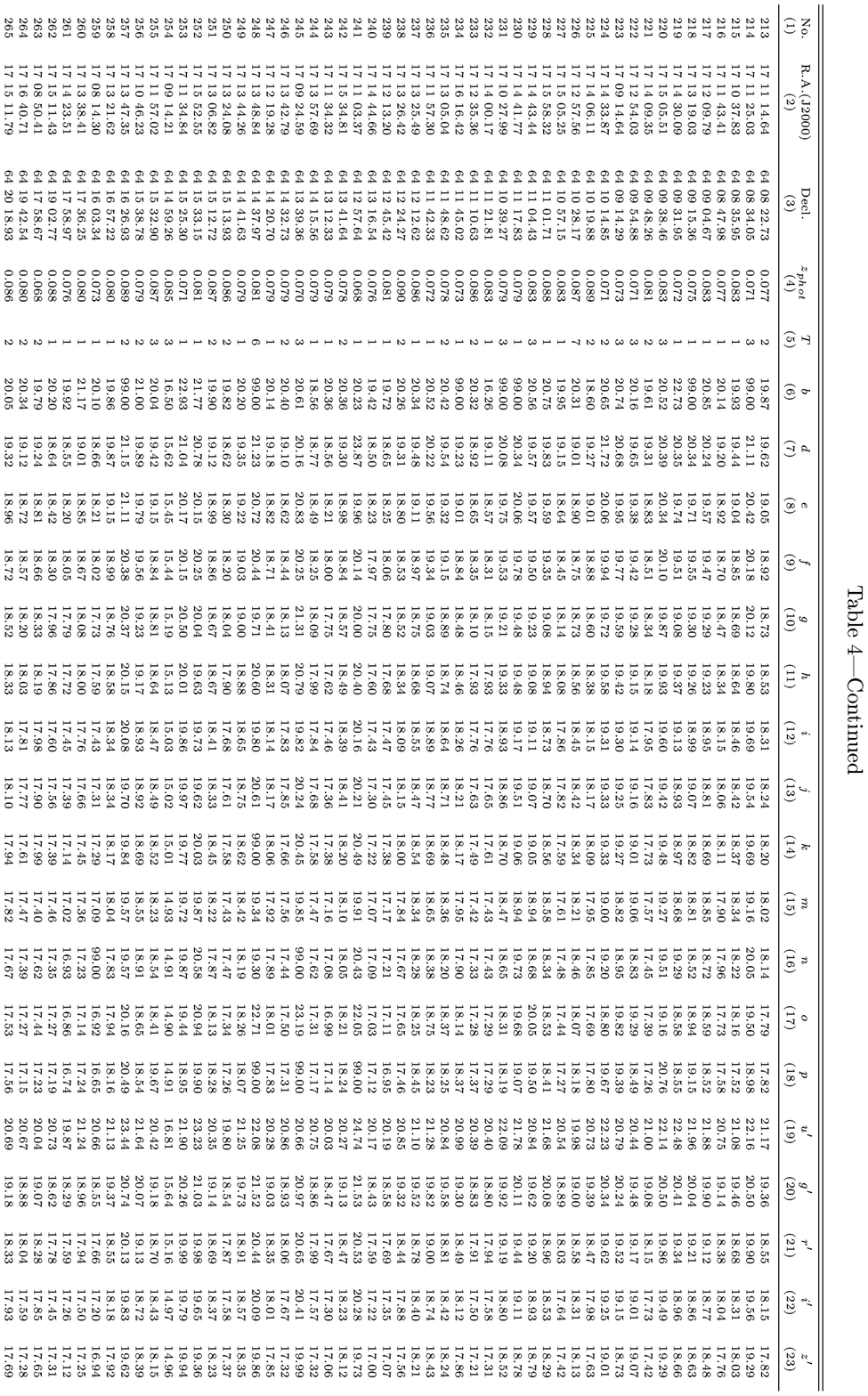}
\end{figure}  \clearpage
\begin{figure}   \epsscale{1.0}
\plotone{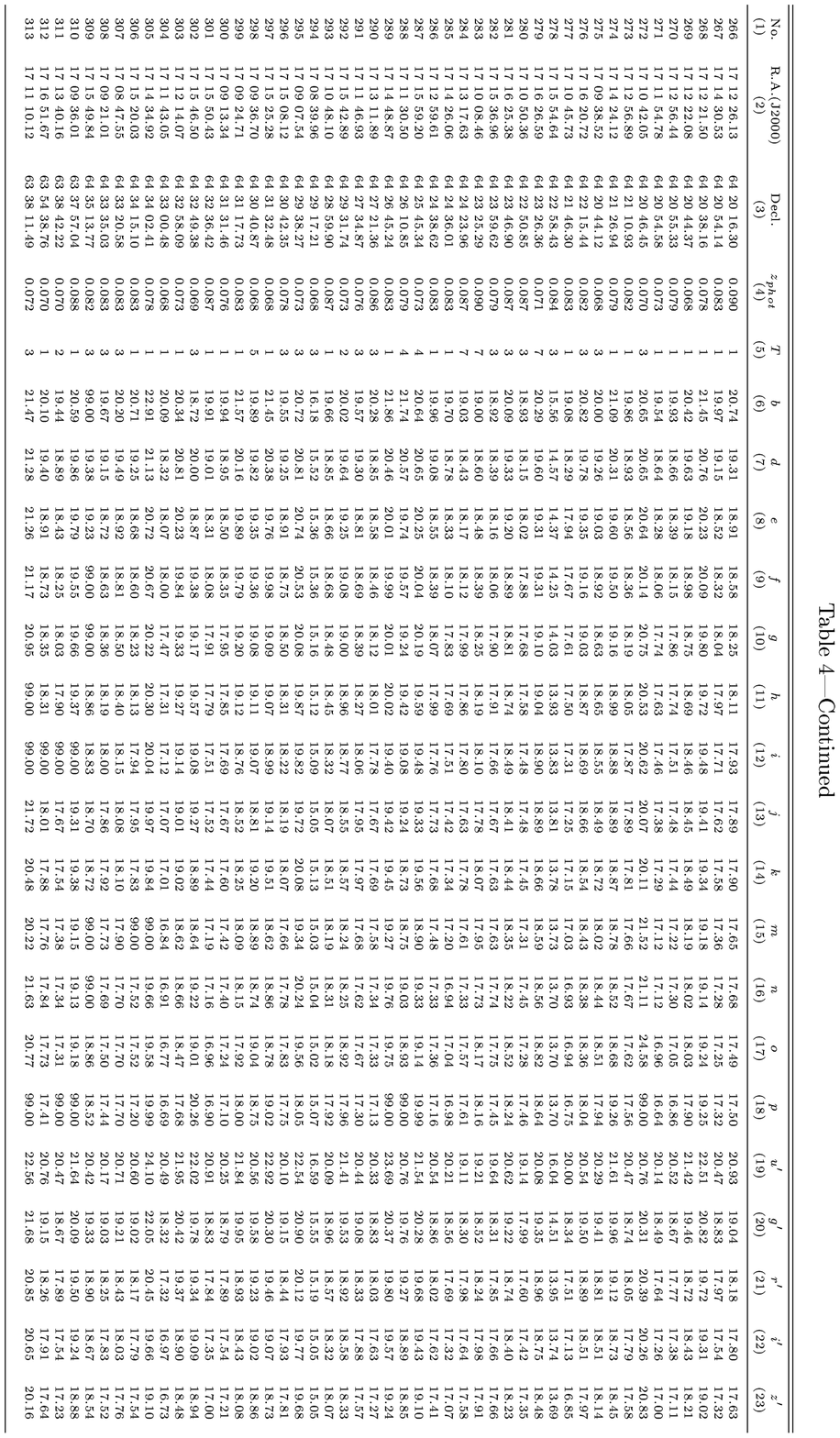}
\end{figure} \clearpage
%\begin{figure} \epsscale{1.0}
%\plotone{tab4_7.eps}
%\end{figure}  \clearpage
%\begin{figure}   \epsscale{1.0}
%\plotone{tab4_8.eps}
%\end{figure}
\end{document}